\documentclass[12pt]{article}

\usepackage{graphicx}
\usepackage{amsmath,amsfonts}
\usepackage{mathtools}
\usepackage{tikz}
\usepackage{booktabs}
\usepackage{multirow}
\usepackage{tabularx}
\usepackage{soul}
\usepackage{fancyhdr}
\usepackage{paralist}
\usepackage{setspace}
\usepackage{subcaption}
\usepackage{epstopdf}
\usepackage{authblk}
\usepackage[UKenglish]{babel}
\usepackage[UKenglish]{isodate}
\usepackage[round]{natbib}
\usepackage{csquotes}
\usepackage[title]{appendix}
\usepackage{floatrow}
\usepackage[section]{placeins} 
\usepackage[T1]{fontenc}
\usepackage{floatrow} 
\usepackage[title]{appendix}
\usepackage{caption}
\usepackage{textcomp}
\usepackage[margin=1in]{geometry}

\usepackage{xr}
\definecolor{orange2}{RGB}{255,185,0}

\usepackage{titlesec}
\titlelabel{\thetitle \quad}
\usepackage{hyperref}
\hypersetup{
     colorlinks = true,
     allcolors = blue,
 }

\newcounter{algorithm}

\newcolumntype{Y}{>{\arraybackslash}X}
\newcolumntype{Z}{>{\centering \arraybackslash}X}

\newcommand\E{\mathbb{E}} 
\newcommand{\defeq}{\vcentcolon=}

\newcommand\blfootnote[1]{%
	\begingroup
	\renewcommand\thefootnote{}\footnote{#1}%
	\addtocounter{footnote}{-1}%
	\endgroup
}

\title{\vspace{-2cm}A Model of the Fed's View on Inflation}
\date{First draft: December 2017}

\author[1]{\normalsize Thomas Hasenzagl}
\author[2]{Filippo Pellegrino}
\author[3]{Lucrezia Reichlin}
\author[4]{Giovanni Ricco}
\affil[1]{\it\small \vspace{-0.3cm}University of Minnesota}
\affil[2]{\it\small \vspace{-0.3cm} London School of Economics and Political Science}
\affil[3]{\it\small \vspace{-0.3cm} London Business School, Now-Casting Economics, and CEPR}
\affil[4]{\it\small University of Warwick, OFCE-SciencesPo, and CEPR}

\begin{document}
\doublespacing

\maketitle
\blfootnote{We thank Filippo Altissimo, Guido Ascari, Travis Berge, Nuno Coimbra, Davide Debortoli, Marco del Negro, Romain Faquet, Yuriy Gorodnichenko, Rodrigo Guimaraes, Michael McMahon, Elmar Mertens, Geert Mesters, Frank Packer, Ricardo Reis, Barbara Rossi, Francesco Zanetti, and conference participants at 2019 NBER-NSF SBIES, 2017 EEA Conference, ECB RCC5 Mini-Workshop on `How to treat trends in macro-econometrics', 1st Vienna Workshop on Economic Forecasting, 2018 ABFER conference, 2018 IAAE Annual Conference, IRTG - Summer Camp 2018, Oxford NuCAMP 2018, NBP-EABCN Conference `Challenges in Understanding the Monetary Transmission Mechanism', 3$^\text{rd}$ Research Conference MMCN, EABCN - Conference on `Advances in Business Cycle Analysis', 2020 Cleveland Fed/ECB inflation conference, and UPF, PSE, OFCE-SciencesPo, Banque de France, S\'eminaire Fourgeaud at the French DG Tr\'esor seminars for helpful comments and suggestions.}
\thispagestyle{empty}

\vspace{-1.5cm}\begin{abstract}
\begin{singlespace}
		We develop a medium-size semi-structural time series model of inflation dynamics that is consistent with the view -- often expressed by central banks -- that three components are important: a trend anchored by long-run expectations, a Phillips curve and temporary fluctuations in energy prices. We find that a stable long-term inflation trend and a well identified steep Phillips curve are consistent with the data, but they imply potential output declining since the new millennium and energy prices affecting headline inflation not only via the Phillips curve but also via an independent expectational channel. A high-frequency energy price cycle can be related to global factors affecting the commodity market, and often overpowers the Phillips curve thereby explaining the inflation puzzles of the last ten years. 
\end{singlespace}
\end{abstract}
\vspace{0.5cm}

\begin{singlespace}\small{\textbf{Keywords:} Phillips curve, inflation dynamics, output gap, Okun's law, unobserved components, Bayesian estimation.}\end{singlespace}

\newpage
\setstretch{1.5}
\normalsize

\begin{quote}
	Inflation is characterized by an underlying trend that has been essentially constant since the mid-1990s; [\ldots]. Theory and evidence suggest that this trend is strongly influenced by inflation expectations that, in turn, depend on monetary policy. In particular, the remarkable stability of various measures of expected inflation in recent years presumably represents the fruits of the Federal Reserve's sustained effort since the early 1980s to bring down and stabilize inflation at a low level. The anchoring of inflation expectations [\ldots] does not, however, prevent actual inflation from fluctuating from year to year in response to the temporary influence of movements in energy prices and other disturbances. In addition, inflation will tend to run above or below its underlying trend to the extent that resource utilization -- which may serve as an indicator of firms' marginal costs -- is persistently high or low.
	
	\begin{flushright}                                      
		\cite{Yellen2016}, `\href{https://www.federalreserve.gov/newsevents/speech/yellen20161014a.htm}{Macroeconomic Research After the Crisis}'\\
		Speech for the 60th Boston Fed Conference
	\end{flushright}                                      
\end{quote}

The quote by Janet Yellen reflects a view, widely shared by policy makers and central bankers, which maintains that three components matter for inflation dynamics: trend-expectations, oil prices, and the degree of resource utilisation in the economy. Similarly, most macroeconomic modelling is based on these three core ideas:  some measure of slack affects short term fluctuations of inflation via a Phillips curve; monetary policy, via expectations, shapes its long run trend; and oil price and other idiosyncratic shocks explain the volatile component of headline inflation. While models that incorporate these ideas use a variety of different auxiliary assumptions (for example on the nature of expectations, the functional form of key equations, and the channels of propagation of the shocks) these three components remain the building blocks of a shared narrative. In this paper, we call this broadly and loosely defined understanding of inflation dynamics the `Fed's view'.

Recent empirical evidence has challenged this view. Indeed, the  literature presents a wide range of contrasting findings, including on the existence, stability, and steepness of the slope of the Phillips curve, and regarding the degree of anchoring of inflation expectations.\footnote{A survey of the extensive empirical literature on the PC is beyond the scope of this paper. For a recent survey of the New Keynesian Phillips curve focussing on univariate limited-information methods, see \cite{Mavroeidisetal} For a review of results using full-information methods to estimate dynamic stochastic general equilibrium (DSGE) models, see  \cite{doi:10.1080/07474930701220071}. \cite{RePEc:anr:reveco:v:5:y:2013:p:133-163} review the use  of microeconomic data to study price dynamics. \cite{RePEc:nbr:nberwo:23304} discuss the incorporation of survey data on inflation expectations in models of inflation dynamics. Other surveys, providing complementary approaches, include \cite{OBES:OBES094}, \cite{RePEc:ice:wpaper:wp31}, \cite{JMCB:JMCB019}, \cite{JAE:JAE1011}, \cite{Gordon2011}, and \cite{RePEc:bla:jecsur:v:25:y:2011:i:4:p:737-768}.}  First, many studies have found the Phillips curve to be unstable, hard to identify,  and weak  or disappearing in recent samples (see results and discussions in \citealp{IMFdog}, \citealp{RePEc:bin:bpeajo:v:42:y:2011:i:2011-01:p:337-405}, \citealp{RePEc:iie:wpaper:wp15-19} and \citealp{RePEc:cfm:wpaper:1815}). Second, Phillips curve based forecasting models have been shown to perform poorly with respect to naive benchmarks, pointing to the irrelevance of slack measures for explaining inflation dynamics  (see, \citealp{ATKE}, \citealp{stock2007has, RePEc:fip:fedbcp:y:2008:n:53:x:2}, and also \citealp{RePEc:fip:fedpwp:11-40}, \citealp{RePEc:cpr:ceprdp:11925},  and \citealp{RePEc:cpr:ceprdp:12652} for recent evidence and relevant discussion).  Third, a small but increasingly important literature has challenged the idea that expectations are fully anchored and forward-looking. For example, papers have connected the `missing disinflation puzzle' of the post-2008 crisis period to the partial disanchoring of consumers' inflation expectations that, in turn, can be accounted for by the evolution of oil prices (see \citealp{COIBGORO}, and \citealp{RePEc:nbr:nberwo:23304}). 

This paper revisits some of the evidence on the reduced form Phillips curve, in the spirit of \cite{ECCA:ECCA283}, by assessing the Fed's view of inflation dynamics through the lens of a stylised statistical model that is informed by economic theory and incorporates economic expectations while allowing for deviations from perfect information and full rationality. Our modelling strategy can be defined as `semi-structural' since it incorporates minimal identifying assumptions from a general class of economic models, but lets the data speak on key aspects, such as expectation formation, the nature of the Phillips curve, and the  role of oil prices.  In this sense it occupies the middle ground between a fully specified Dynamic Stochastic General Equilibrium (DSGE) model and  a Vector Auto Regressive (VAR) model.

Our specification in reduced form is compatible with and nests several potentially different forward- and backward-looking structural Phillips curve models, including the standard New-Keynesian Phillips curve (NKPC), in which inflation is a purely forward-looking process, driven by expectations of future real economic activity. Moreover, the model  allows survey data on agents' expectations on inflation to depart from the full-information rational expectations benchmark without imposing any specific form of information frictions. We do not require either of the two surveys to be an efficient and unbiased predictor of future inflation and allow for temporary and permanent deviations from a rational forecast, potentially capturing measurement and observational errors, as well as a time-dependent bias in inflation expectations. 

A key feature of the approach is the modelling of oil prices and the different channels through which energy prices can affect inflation. One way is through production marginal costs and the Phillips curve -- oil prices can affect the business cycle component and hence co-determine  the output gap.\footnote{A large and important literature has analysed the connection between demand and supply oil shocks and the business cycles (see, for example, \citealp{RePEc:bin:bpeajo:v:47:y:2016:i:2016-02:p:287-357}, \citealp{RePEc:elg:eechap:14429_1}, \citealp{RePEc:wly:jmoncb:v:49:y:2017:i:8:p:1747-1776}).} Furthermore, in the model, oil disturbances can affect headline prices directly  via energy services, which are part of the consumption basket, but also potentially via expectation formation, in line with the findings of \cite{COIBGORO}. These two channels are captured by studying the differential impact of a second cycle, that we label `energy price cycle', on headline and core inflation. The energy price cycle captures the potential common dynamics between oil prices, inflation expectations, and inflation but it does not affect the domestic output gap and the real variables.\footnote{It is important to stress that our assumption of an energy price cycle orthogonal to the business cycle and not affecting the real variables should not be seen as literally present in the data structure. It is a convenient statistical device which helps teasing out components in the price dynamics, at higher frequencies than those of the standard business cycle, and that can have weak or negligible impact on the US output gap and labour market.}

In an extension of the model which includes proxies of global economic activity we analyse  whether the energy price cycle  reflects global demand and the commodity price cycle. Our results suggest that  the energy price cycle is associated with oil supply shocks and financial shocks in the commodity markets rather than global demand.

Inflation is modelled as being driven by three components: (i) long term inflation expectations; (ii) a  stationary stochastic cycle, which captures multivariate and lagged commonalities in real, nominal (including energy prices) and labour market variables at business cycle frequencies. This cycle connects the output gap to prices and their expectations via a Phillips curve relationship and to unemployment via the Okun's law; (iii)  a stationary stochastic cycle capturing the common dynamics between oil prices, inflation expectations, and CPI inflation but not affecting real variables. The model also identifies other key economic objects such as output potential, trend employment, and equilibrium unemployment, in the form of unit root trends.



Results suggest  that the Phillips curve is alive and well and has been fairly stable since the early 1980s.\footnote{While we observe that a fixed parameter model is able to capture a stable Phillips curve from the 1980s, it is possible that time-variation in the parameters or stochastic volatility may be important over a longer sample \citep[see][]{stock2007has,RePEc:een:camaaa:2017-60}. We do not explore this possibility in this paper. Indeed, estimation uncertainty is likely to obfuscate all gains coming from a more sophisticated model.} Importantly, our cycle decomposition shows that the business cycle is not always the dominant component. Large oil price fluctuations can move prices away from the real-nominal relationship both by directly impacting energy services prices and by shifting  consumers' expectations away from the rational forecast -- `disanchoring' them -- and hence inducing expectation driven fluctuations in prices. This result confirms the intuition of \cite{COIBGORO}. We  provide confirmation of the importance of using expectational data to identify both trend inflation and the Phillips curve, while dealing with disturbances to expectations that, albeit reflected in inflation, are unrelated to real variables and fundamentals. From a policy perspective, the stable inflation trend is an indication of the Fed's success in anchoring expectations. However, our results also point to the challenges that policymakers have to overcome in guiding expectations and stabilising the economy in the presence of large energy price disturbances.  
 
There are several by-products of our analysis:  we obtain a model-consistent estimate of the output gap and potential output; we also assess the stability of Okun's law and the quality of core inflation as an indicator of underlying inflation. Indeed, our approach generates an indicator of cyclical inflation which is clean not only from the direct effect of oil prices, as is the case for core inflation, but also from their indirect effects.

The paper starts with a brief discussion of our methodology and related literature, in the remainder of this section. In \autoref{sec:model_strategy} we then introduce a stylised model of inflation dynamics which provides the intuition for our approach. In \autoref{sec:model} and \ref{sec:model_estimation} we specify the empirical model while in \autoref{sec:empirics} and \ref{sec:global} we discuss empirical results. In \autoref{sec:forecasting} we present an out-of-sample  forecasting evaluation and the last section concludes.


{\bf Contribution and Related Literature.} From the statistical point of view, the model has a number of attractive features: it does not rely on arbitrary preliminary detrending of the data which may create distortions, it contains a rich lag structure allowing us to capture dynamic heterogeneity amongst variables, it allows us to perform conjunctural analysis and historical decompositions of variables into cyclical and trend components, and it is sufficiently efficient and parsimonious to be used as a forecasting tool.  The unit root trend common to inflation and inflation forecasts can be related to agents' long-term expectations, under the assumption that the `law of iterated expectations' holds (see \citealp{beveridge1981new} and \citealp{RePEc:tpr:restat:v:98:y:2016:i:5:p:950-967}). In fact, the impact of all transitory components has to be zero in the long run.\footnote{A discussion on the conditions under which survey data can be employed to study the PC is in \cite{RePEc:bla:ecinqu:v:49:y:2011:i:1:p:13-25}.} 

Our econometric representation is general in the sense described but has a structure that is  motivated by the objective of parsimony. Indeed, our model  can be understood as a restricted VAR  model where, by adopting minimal economic restrictions to identify the potentially different dynamic components of inflation, we induce `informed' parsimony thereby helping with signal extraction and forecasting. The proposed decomposition leads to a rather complex state space form. In order to deal with this complexity, we estimate the model using Bayesian methods. A Bayesian approach in the context of a similar but simpler model has been proposed by  \cite{10.2307/27638958} who implement  a Bayesian version of the work of \cite{RePEc:bes:jnlbes:v:12:y:1994:i:3:p:361-68}, by \cite{doi:10.1111/jmcb.12388} who propose a Bayesian model comparison focussing on trend-cycle decompositions of output and, more recently, by \cite{RePEc:ecb:ecbwps:20161966}. The latter paper is the closest to our work but focuses on estimating measures of the  output gap in the Euro Area rather than on providing a decomposition that can be used for studying the drivers of inflation dynamics.  Our paper also shares a similar approach and methodology with \cite{RePEc:bin:bpeajo:v:48:y:2017:i:2017-01:p:235-316}, who employ a flexible VAR model that incorporates long-term survey expectations, to estimate common trends and study the natural rate of interest in the US.

Our work builds on the tradition of structural time series models \citep[see][]{harvey1985trends}, where observed time series are modelled as the sum of unobserved components: common and idiosyncratic trends, and cycles. In doing this, and by focussing on inflation dynamics, this paper relates to the literature on the output gap, the Phillips curve, and trend inflation estimation with unobserved components models, started by \cite{RePEc:bes:jnlbes:v:12:y:1994:i:3:p:361-68}. Similarly to \cite{JAE:JAE2411} and \cite{RePEc:ecb:ecbwps:20161966}, we do not pre-filter data to stationarity, but model their low frequency behaviour by allowing for trends. As in {\cite{f90dccd72233401bb8488cae6850a509} and \cite{RePEc:tpr:restat:v:90:y:2008:i:4:p:805-811},  we use multiple real activity indicators to increase the reliability of the output gap estimates. Also, our work relates to a number of papers which have studied trend inflation in unobserved component models augmented with data on medium-/long-term inflation expectations, as for example, \cite{RePEc:eee:intfor:v:30:y:2014:i:3:p:426-448}, and \cite{RePEc:tpr:restat:v:98:y:2016:i:5:p:950-967}.

\section{A Stylised Model for Inflation Dynamics}\label{sec:model_strategy}

At the core of our empirical approach lies a stylised full information rational expectations model for inflation and output. In this section we discuss the intuition and basic building blocks. We assume that inflation and output can be decomposed into three components: (i) independent trends determining output potential $\mu_t^{y}$ and trend inflation $\mu_t^{\pi}$; (ii) a common stationary cycle relating nominal and real variables (the output cycle is interpreted as  the output gap) $\widehat \psi_t$; and (iii) some independent (white noise) disturbances to output and inflation, $\psi^{y}_t$ and $\psi^{\pi}_t$, that can be thought of as classic measurement error or idiosyncratic shocks. We have:
\begin{eqnarray}
y_t &=& \mu_t^{y} + \widehat \psi_t + \psi^{y}_t \ ,\label{NKmodel1} \\
\pi_t &=& \mu_t^{\pi} + \delta_\pi \widehat \psi_t + \psi^{\pi}_t  \ ,\label{NKmodel2}
\end{eqnarray}
where the independent trends are assumed to be unit-root processes (with a drift in output)
\begin{eqnarray}
\mu_t^{y} &=& \mu_0 + \mu_{t-1}^{y} + u_t^y \ , \\
\mu_t^{\pi} &=&  \mu_{t-1}^{\pi} + u_t^\pi \ .
\end{eqnarray}

The economic interpretation of the different trend and cycle components is standard \citep[see, for example the discussion in][]{RePEc:fip:fedgsq:863}. The output trend  -- i.e. the output potential, capturing the long-term growth of the economy -- is usually thought of as driven by technological innovation. Inflation fluctuates around a longer-term trend that, at least in recent times, has been essentially stable. Theory relates this trend inflation to inflation expectations that, in turn, are shaped by the conduct of monetary policy -- for example, by policymakers' targets. Shocks of a different nature can impact marginal production costs and modify the intensity of resource utilisation in the economy, thus, temporarily pushing output away from its balanced growth path. The shortfall of actual GDP from potential output is the output gap $\widehat \psi_t$. The slack in the economy is reflected in the short-run cyclical fluctuations of inflation around its trend, in the presence of price rigidity. This relationship is generally  described by an expectations-augmented Phillips curve in theoretical models. Finally, a nontrivial fraction of the quarter-to-quarter variability of inflation and output is attributable to independent and idiosyncratic shocks.

In line with the econometric literature on the output gap, we assume that $\widehat \psi_t$ is a stationary process with stochastic cyclical behaviour. The simplest process allowing for such a stochastic cycle is an AR(2) process with complex roots of the form
\begin{equation}
\widehat \psi_t = \alpha_1 \widehat \psi_{t-1} + \alpha_2 \widehat \psi_{t-2} +  v_{t} \ .
\end{equation}
Indeed, the AR(2) model can be written in a different and slightly more general form, displaying its pseudo-cyclical behaviour more clearly , i.e.
\begin{align}
    &\widehat \psi_{t} = \rho \cos(\lambda) \widehat \psi_{t-1} + \rho \sin(\lambda) \widehat \psi^*_{t-1} + v_{t} \ , \label{output_gap1} \\
    &\widehat \psi^*_{t} = -\rho \sin(\lambda) \widehat \psi_{t-1} + \rho \cos(\lambda) \widehat \psi^*_{t-1} + v^*_{t}  \nonumber \ ,
\end{align}  
where the parameters $0 \leq \lambda \leq  \pi$ and $0 \leq \rho \leq 1$ can be interpreted, respectively,  as the frequency of the cycle and the damping factor on the amplitude while $\widehat \psi^*_{t}$ is a modelling auxiliary cycle  and $v_{t}$ and $v^*_{t}$ are uncorrelated white noise disturbances \citep[see][]{harvey1990forecasting}.\footnote{It is straightforward to show that the model can be rewritten as
$$
(1-2\rho \cos(\lambda)L + \rho^2 L^2) \widehat \psi_{t} = (1 - \rho \cos(\lambda)L)v_{t} + (\rho \sin(\lambda)L)v^*_{t} \ .
$$
Hence, under the restriction $\sigma^2_{v} = 0$, the solution of the model is an AR(2), otherwise an ARMA(2,1). The intuition for the use of the auxiliary cycle is closely related to the standard multivariate AR(1) representation of univariate AR(p) processes.} The disturbances make the cycle stochastic rather than deterministic and, if $\rho < 1$, the process is stationary.

By assuming an output gap that is a stationary solution to an AR(2) process, the model in Eq. (\ref{NKmodel1}-\ref{NKmodel2}) admits a hybrid expectations-augmented New Keynesian Phillips Curve connecting the cyclical components of output, inflation, and inflation expectations, of the form
\begin{equation}
\widehat \pi_t = \sum_{i=1}^2 \delta_i \widehat \pi_{t-i}  + \beta \mathbb{E}_t \left[\widehat \pi_{t+1} \right] + \kappa \widehat y_t  +  \varepsilon_{t} \ ,
\end{equation}
where hats indicate deviations from trends.\footnote{Empirical studies often feature hybrid Phillips curves to account for inflation persistence \citep[a recent survey is in][]{RePEc:bla:jecsur:v:25:y:2011:i:4:p:737-768}. Several different mechanisms have been proposed in the literature to introduce hybrid Phillips curves such as indexation assumptions \citep[e.g.][]{RePEc:eee:moneco:v:44:y:1999:i:2:p:195-222}, state-contingent pricing \citep[e.g.][]{RePEc:oup:qjecon:v:114:y:1999:i:2:p:655-690}, or deviations from rational expectations assumption \citep[e.g.][]{RePEc:eee:moneco:v:50:y:2003:i:4:p:915-944, RePEc:eee:moneco:v:54:y:2007:i:7:p:2065-2082}.} In this model, rational expectations agents correctly form model-consistent expectations about inflation, that is 
\begin{eqnarray}
\mathbb{E}_t \left[\pi_{t+1} \right] = \mathbb{E}_t \left[\mu_{t+1}^{\pi} + \delta_\pi \widehat \psi_{t+1} + \psi_{t+1}^{\pi}\right] &=& \mu_t^{\pi} + \delta_\pi  (\alpha_1 \widehat \psi_{t} + \alpha_2 \widehat \psi_{t-1})  \nonumber \\
&=& \mu_t^{\pi} + \delta_{exp,1} \widehat \psi_{t} +  \delta_{exp,2} \widehat\psi_{t-1} \ . \nonumber
\end{eqnarray}
The model can be written, in a compact reduced form representation in terms of the common cycle, the trend common to inflation and inflation expectations, and the trend capturing output potential (as well as the idiosyncratic disturbances):
\begin{align}\label{REE_PC}    
    \hspace*{-0.8cm}
    \begin{pmatrix}
        y_{t} \\ \pi_{t} \\ \mathbb{E}_t \left[\pi_{t+1} \right]
    \end{pmatrix}=  
    \begin{pmatrix}
        1 & 0  \\ \delta_{\pi} & 1 \\  \delta_{exp,1} +  \delta_{exp,2}L &1 
    \end{pmatrix}    
    \begin{pmatrix}
        \widehat \psi_{t} \\ \mu^{\pi}_{t}
    \end{pmatrix}+    
    \begin{pmatrix}
       \mu_{t}^{y} \\ 0 \\ 0
    \end{pmatrix}
    +    
    \begin{pmatrix}
        \psi_{t}^{y} \\ \psi_{t}^{\pi} \\ 0
    \end{pmatrix} \ .
\end{align}    
In principle, this simple set of equations can also accommodate different specifications for the Phillips Curve, under suitable parameter restrictions. For example, an AR(1) $\widehat \psi_t$ would be the solution to a purely forward looking New-Keynesian Phillips Curve.  It also nests the backwards looking `Old-Keynesian' Phillips curve connecting output gap and prices -- as in the `triangle model of inflation' \citep[see][]{f90dccd72233401bb8488cae6850a509, Gordon1990}.

Also, in line with the interpretation proposed, it is worth noting that trend inflation corresponds to the long-run forecast for inflation, which implies
\begin{equation}
\lim_{h\to\infty}\E_t[\pi_{t+h}] = \mu_t^\pi \ ,
\end{equation}
in the spirit of \cite{beveridge1981new}, and that trend output informs expectations of growth in the long run:
\begin{equation}
\lim_{h\to\infty}\E_t[ y_{t+h}] = \lim_{h\to\infty} \{\mu_0h + \mu_t^y\} \ .
\end{equation}

While such a stylised rational expectations model can provide the gist of the intuition for our econometric model, it is likely to be too simple as an empirical representation of business cycle dynamics.\footnote{An estimated version of this model provides an unsatisfactory representation of the structure of the data. Results are available in the Online Appendix.} First, it does not allow for dynamic heterogeneity, and hence nominal and real variables fluctuate only as contemporaneously connected by the slack in the economy, in contrast with the evidence that  prices and labour market variables  respond with lags to the slack in production. In fact, output is linked to unemployment via Okun's law and to inflation via the Phillips curve relationship which may involve lagging dynamics. These fundamental relationships connect potentially different measures of the slack in the economy, such as the output gap and the cyclical component of unemployment -- i.e. the difference between the unemployment rate and its normal long-run level (equilibrium unemployment)\footnote{For example, the measure of slack that is adopted in policy analysis by the Fed is obtained as the difference between the unemployment rate and the Congressional Budget Office's (CBO) historical series for the long-run natural rate \citep[as in][]{RePEc:fip:fedgsq:863}.} -- and inform fluctuations at business cycle frequency in other real and nominal variables. 

Second, in modelling price dynamics, forecasters and policymakers often distinguish between changes in energy and food prices -- which enter into headline inflation -- and movements in the prices of other goods and services -- that is, core inflation.\footnote{The price index for total consumer price (headline) inflation $\pi_t$  is decomposed as
\begin{equation}
\pi_t = \pi_t^c + \upsilon_1 \pi_t^{en} +  \upsilon_2 \pi_t^{food} \ ,
\end{equation}
where $\pi_t^c$ is core CPI inflation, and $\pi_t^{en}$ and $\pi_t^{food}$ are, respectively, the growth rate for prices of consumer energy goods and services and prices of food, both expressed relative to core CPI prices; and  $\upsilon_1$, and  $\upsilon_2$ are the weights of energy and food in total consumption. In the rest of the paper we focus on the energy price component and abstract from food prices. Interestingly, both commodities are subject to the effect of global factors and a few papers have reported a substantial share of co-movement between energy and food prices \citep[see, for example,][]{RePEc:bla:ecpoli:v:29:y:2014:i:80:p:691-747}.} This is because food and energy prices tend to be extremely volatile and  influenced by factors that are disconnected from the slack in the economy and that are beyond the control of monetary policy. Examples are  international political events -- as is the case for oil price -- as well as weather or diseases -- as for food and beverages.\footnote{While the Federal Reserve's inflation objective is defined in terms of the overall change in consumer prices, core inflation is considered to provide a better indicator than total inflation for the developments in prices, in the medium term.}  This decomposition is important to study how slack in real output is transmitted to prices, by separating the direct impact of energy price shocks onto energy products, from their role as cost push shocks in production. 

Finally, it has been argued in the literature that, once inflation expectations are admitted to a forward- or backward-looking Phillips curve equation, it is also possible that economic disturbances impact prices without any intermediating transmission through the output gap or other measures of slack in the economy \citep[see, for example,][]{RePEc:fip:fedbcp:y:2008:n:53:x:7}. In this spirit, \cite{COIBGORO}  argue that the absence of disinflation during the Great Recession  can be explained by the rise of consumers' inflation expectations between 2009 and 2011 due to the increase in oil prices in this period.  Also, while macro-variables are likely to be affected by non-classical measurement error, agents' expectations, as captured by consumers' and professional forecasters' surveys, are likely to be only partially in line with national accounting definitions of aggregate prices and can introduce measurement errors and biases of a different nature.\footnote{For example, especially in consumer surveys the forecast horizon may be loosely defined  while the relevant price index may be left unspecified. Also, projections are often reported at different frequencies and can have different forecasting points.} 

In the next section, we present an empirical model that expands on the core model to accommodate these possibly important aspects of business cycle and inflation dynamics.

\section{An Empirical Trend-Cycle Model}\label{sec:model}

Our benchmark empirical model expands on the core rational expectations model presented in the previous section to incorporate a rich information set including output, employment, and the unemployment rate -- as measures of real activity and labor market  developments --, CPI inflation, core CPI inflation and  consumers' and professionals' forecasts for  one year ahead inflation -- as proxies for economic agents' inflation expectations --, and oil prices to proxy for energy prices. To capture the complex dynamics relationships among the variables, we generalise the stylised model presented in the previous section by incorporating dynamic heterogeneity in the relationship linking real variables, labour market outcomes, and prices and by allowing for deviations from perfect rationality. 

Our model provides an empirical specification of a number of key macroeconomic concepts. A unit root trend with drift provides a time varying measure of output potential, while the trend in employment/unemployment captures the evolution of equilibrium unemployment. The cyclical component of unemployment connects to fluctuations in output at business cycle frequency via an Okun's law that involves the output gap and its lagged value. This allows business cycle fluctuation to have dynamic heterogeneity and the labour market to respond with a lag to the slack in the economy. A unit root trend -- common to headline and core CPI inflation, and inflation expectations -- captures the inflation trend shaping long term expectations. The slack in the economy is reflected in the short-run cyclical fluctuations of inflation (and expectations) via a Phillips curve relationship involving the output gap and its lagged value that accommodates for a slow adjustment of prices to slack, in the presence of nominal rigidities. Also, oil prices are allowed to co-move along the business cycle  and possibly its lagged value, due to demand effects or mark-up shocks. The fact that the cyclical component of output informs economy-wide lead-lag fluctuations in both labour market and nominal variables supports the interpretation of the output gap as a measure of the business cycle.

\begin{table}[t!]
    \caption{Data and transformations}
    \label{Tab:data}    
	\small
	\begin{tabularx}{\textwidth}{@{}lZZZ@{}}
		\toprule
	    Variable &  Symbol & Mnemonic & Transformation \\
		\midrule
		Real GDP & $y_t$ & $y$ & Levels  \\
		Employment & $e_t$ & $e$ & Levels \\
		Unemployment rate & $u_t$ & $u$ & Levels \\
		Oil price & $oil_t$ & $oil$ & Levels \\
		CPI inflation & $\pi_t$ & $\pi$ & YoY \\                    
        		Core CPI inflation & $\pi^{c}_t$ & $\pi^{c}$ & YoY \\       
		UoM: Expected inflation & $F_{t}^{uom}\pi_{t+4}$ & $uom$ & Levels \\
		SPF: Expected CPI & $F_{t}^{spf}\pi_{t+4}$ & $spf$ & Levels \\
		\bottomrule
	\end{tabularx} 
    \floatfoot{\textbf{Note:} The table lists the macroeconomic variables used in the empirical model. `UoM: Expected inflation' is the University of Michigan, 12-months ahead expected inflation rate. `SPF: Expected CPI' is the Survey of Professional Forecasters, 4-quarters ahead expected CPI inflation rate. The oil price is the West Texas Intermediate Spot oil price.}
\end{table}

We also design the model to be able to account for several potential deviations from the rational expectations benchmark. In particular, we allow for (i) oil price disturbances to affect prices either directly via energy prices in headline CPI, or via economic agents' forecasts by inducing a transitory disanchoring of expectations, with a stationary cycle connecting oil prices, expectations, and inflation but not the measure of slack in the economy; (ii) a time varying bias i.e. a permanent disanchoring of expectations in the form of unit root processes; (iii) non-classic measurement error in the variables and other sources of coloured noise.

We summarise these modelling choices in the following assumptions.
\begin{itemize}
\item[\bf Assumption 1] CPI headline inflation, core CPI inflation and agents' inflation expectations (consumers' and professional forecasters') share a common random walk trend (viz. {\bf  trend inflation}). 
\item[\bf Assumption 2] Real output, employment, and unemployment have independent trends modelled with unit roots, with a drift for output and employment (i.e. {\bf potential output} and {\bf equilibrium employment/unemployment} respectively).
\item[\bf Assumption 3] Business cycle fluctuations in output are described by a stationary process with stochastic cycle in the form of an ARMA(2,1) process with complex roots (i.e. {\bf output gap}).
\item[\bf Assumption 4] Inflation, inflation expectations, and output are connected by a {\bf Phillips curve} relationship defined as a moving average of the output gap and its first lag. 
\item[\bf Assumption 5] Labour market variables are linked to output via the {\bf Okun's Law} defined as a moving average of the output gap and its first lag. 
\item[\bf Assumption 6] Oil prices co-move with the business cycle via a a moving average of the output gap and its firs lag ({\bf business cycle component of oil prices}).
\item[\bf Assumption 7] Inflation expectations and inflation are connected, via a moving average of order one, to an ARMA(2,1) cycle in oil prices ({\bf Energy cycle}).
\item[\bf Assumption 8] All variables can have an idiosyncratic ARMA(2,1) cycle component, possibly capturing {\bf non-classic measurement error}, {\bf differences in definitions} and other {\bf sources of noise}.
\item[\bf Assumption 9] Agents' (consumers and professional forecasters) expectations have independent idiosyncratic unit roots without drift, capturing {\bf time varying bias in the forecast}.
\item[\bf Assumption 10] All components are mutually orthogonal. 
\end{itemize}

A key and novel feature of our modelling strategy is to allow the oil prices to affect and be affected by both the standard business cycles and what we define as an energy price cycle. Fluctuations in the latter component are reflected in prices and inflation expectations without affecting output and the labour market. This orthogonality assumption is a convenient statistical device helpful in teasing out components in the price dynamics which have weak or negligible impact on the US output gap and labour market, and that may happen at frequencies different from those of the standard business cycle frequency range. 

For the purpose of this analysis the University of Michigan (UoM) consumer survey and the Federal Reserve Bank of Philadelphia's Survey of Professional Forecasts (SPF) one year ahead inflation forecast  were chosen as proxies for consumers' and professionals' expectations. This because they both have relatively long histories and are available at quarterly frequency. Both of them target CPI inflation, either explicitly as is the case for the SPF or, implicitly, by surveying consumers, as is the case for UoM. For both surveys, we employ the median expected price change in the four quarters following the date of the survey, which is consistent with our use of year-on-year inflation. Data incorporated in the model are at quarterly frequency, with the sample starting in Q1 1984 and ending in Q2 2017. All variables enter the model in levels, except for price variables which are transformed to the year-on-year inflation rate (see \autoref{Tab:data} for details). 

Our model in $x_t \defeq \{y_{t}, \allowbreak e_{t}, \allowbreak u_{t}, oil_{t}, \allowbreak \pi_{t}, \allowbreak \pi^{c}_{t},\allowbreak F_{t}^{uom}\pi_{t+4}, \allowbreak F_{t}^{spf}\pi_{t+4} \}$ can be written as
\begin{align}
\resizebox{0.91\hsize}{!}{%
$
    \begin{pmatrix}
        y_{t} \\ e_{t} \\ u_{t} \\ oil_{t} \\ \pi_{t} \\ \pi^{c}_{t} \\  F_{t}^{uom}\pi_{t+4} \\ F_{t}^{spf}\pi_{t+4}
    \end{pmatrix}= 
    \begin{pmatrix}
        1 & 0 & 0 \\ 
        \delta_{e,1} + \delta_{e,2}L & 0 &0 \\ 
        \delta_{u,1} + \delta_{u,2}L & 0 &0 \\ 
        \delta_{oil,1} + \delta_{oil,2}L & 1 & 0 \\ 
        \delta_{\pi,1} + \delta_{\pi,2}L & \gamma_{\pi,1} + \gamma_{\pi,2}L& \phi_{\pi} \\ 
        \delta_{\pi^{c}, 1} + \delta_{\pi^{c}, 2}L & \gamma_{\pi^{c},1} + \gamma_{\pi^{c},2}L & \phi_{\pi^{c}}\\ 
        \delta_{uom,1} +  \delta_{uom,2}L + \delta_{uom,3}L^2 & \gamma_{uom,1} + \gamma_{uom,2}L & \phi_{uom} \\
        \delta_{spf,1} + \delta_{spf,2}L + \delta_{spf,3}L^2 & \gamma_{spf,1} + \gamma_{spf,2}L & \phi_{spf}   
    \end{pmatrix}    
    \begin{pmatrix}
        \widehat \psi_{t} \\ \psi^{EP}_{t} \\ \mu^{\pi}_{t}
    \end{pmatrix}+    
    \begin{pmatrix}
        \psi_{t}^{y} \\ \psi_{t}^{e} \\ \psi_{t}^{u} \\\psi_{t}^{oil} \\ \psi_{t}^{\pi} \\ \psi_{t}^{\pi^{c}}\\ \psi_{t}^{uom} \\ \psi_{t}^{spf} 
    \end{pmatrix}+  
    \begin{pmatrix}
        \mu_{t}^{y} \\ \mu_{t}^{e} \\ \mu_{t}^{u}\\ \mu_{t}^{oil} \\ 0 \\ 0 \\ \mu_{t}^{uom} \\ \mu_{t}^{spf} 
    \end{pmatrix}
$}
\label{Model}    
\end{align}    
 where $\phi_{\pi}$, $\phi_{\pi^c}$, $\phi_{uom}$, and $\phi_{spf}$ are normalised to have unitary loading of inflation and inflation expectations on trend inflation.\footnote{In the empirical model, the series are standardised so that the standard deviations of their first differences are equal to one. For this reason, we normalise $\phi_{\pi}$, $\phi_{\pi^c}$, $\phi_{uom}$, and $\phi_{spf}$ to the reciprocal of the standard deviation of the first difference of the respective variable.} It is worth noting that our empirical specification in \autoref{Model} would reduce to the stylised rational expectations model in \autoref{REE_PC}, under suitable parametric restrictions. 
In the Online Appendix, we report a number of simplified models and their estimation results to show how different assumptions impact on the final specification of the model. 

Like the output gap in \autoref{output_gap1}, the energy cycle and the idiosyncratic ARMA(2,1) stationary cycles can be written in the following form:
\begin{equation}
    \begin{split}
	\begin{pmatrix}
	    \psi^{j}_{t} \\
	    \psi^{*j}_{t}
	\end{pmatrix} = 
    \rho^{j} 
    \begin{pmatrix}
	    \cos(\lambda^{j}) & \sin(\lambda^{j}) \\
	    -\sin(\lambda^{j}) & \cos(\lambda^{j}) 
	\end{pmatrix}
    \begin{pmatrix}
	    \psi^{j}_{t-1} \\
	    \psi^{*j}_{t-1}
	\end{pmatrix} 
    + 
    \begin{pmatrix}
	    v^{j}_{t} \\
	    v^{*j}_{t}
	\end{pmatrix},
    \quad 
    \begin{pmatrix}
	    v^{j}_{t} \\
	    v^{*j}_{t}
	\end{pmatrix}
    \sim  {\mathcal N}(0, \varsigma^2_{j} \, I_2)
    \end{split}
\label{Cycle}
\end{equation}
where $j \in \{EP, x_{1}, \ldots, x_{n}\}$ and $\psi^{*j}$, as discussed, is a term capturing an auxiliary cycle. For stationarity, we impose $0 < \lambda^{j} \leq \pi$ and  $0 < \rho^{j} < 1$ for all cycles, including the output gap. 

There are four main advantages to modelling the stationary components as restricted ARMA(2,1) processes.  First, this representation nests an AR(2) that is the simplest linear process able of displaying pseudocyclical behaviour of the type it is associated with the business cycle and other economic cycles. Second, it allows for an explicit characterisation of the relevant cyclical parameters -- frequency and decay rate --, over which it is possible to specify transparent priors. Third, it is a very parsimonious representation with a small number of parameters and hence the estimation of many stationary components is computationally feasible. Fourth, the presence of an additional MA(1) component is potentially able to accommodate for additional persistence in the data.

As discussed, the common and idiosyncratic trends are random walks (with/without drifts -- $\mu^j_{0}$) that can be written as
$$
\mu^j_t = \mu^j_{0} + \mu^j_{t-1} + u^j_t, \quad u^j_t \sim {\mathcal N}(0, \sigma^2_j) \ .
$$
All of the stochastic disturbances in the model are assumed to be mutually orthogonal and Gaussian. Finally, it is worth noting that the common and idiosyncratic trends in inflation and inflation expectations are identified up to a constant \citep[see][for a discussion on identification]{bai2015identification}. For the sake of interpretation, we attribute the constant to the common trend so that it is on the same scale as the observed inflation variables.

\section{Bringing the Model to the Data}\label{sec:model_estimation}

\begin{table}[t!]
	\caption{Prior distributions}
	\small
	\begin{tabularx}{\textwidth}{@{}lYYYY@{}}
		\toprule
		Name & Support & Density & Parameter 1 & Parameter 2 \\
		\midrule
		$\delta$, $\gamma$, $\phi$ and $\tau$ & ${\rm I\!R}$ & Normal & 0 & 1000 \\
		$\sigma^2$ and $\varsigma^2$& $(0, \infty)$ & Inverse-Gamma & 3 & 1 \\
		$\rho$ & $[0.001, 0.970]$ & Uniform & 0.001 & 0.970 \\
		$\lambda$ & $[0.001, \pi]$ & Uniform & 0.001 & $\pi$ \\
		\bottomrule
	\end{tabularx}
    \floatfoot{\textbf{Note:} Prior distribution for the model parameters adopted in estimating the model with US data. All of the priors are uniform over the range of the model parameters compatible with our modelling or weakly informative. Boundaries of the uniform priors ensure that the stochastic cycles are stationary and correctly specified according to the restrictions described in \cite{harvey1990forecasting}.}  
	\label{priors}
\end{table}

Our estimation strategy builds on the approach recently suggested by \cite{harvey2007trends}, that adopts modern Bayesian techniques to support the estimation of `structural' trend-cycle models \`a la \cite{harvey1985trends}.  In estimating the model, we elicit prior distributions that are either uniform over the range of the model parameters compatible with our modelling choices (i.e. $0 < \lambda^{j} \leq \pi$ and  $0 < \rho^{j} < 1$), or weakly informative and in the form of very diffuse Normal and Inverse Gamma priors. \autoref{priors} reports the parameters of our prior distributions. 

We maximise and simulate the posterior distributions with a Metropolis-Within-Gibbs algorithm that is structured in two blocks. In the first block, we estimate the state space parameters by the Metropolis algorithm and, in the second block, we use the Gibbs algorithm to draw unobserved states conditional on model parameters. Relevant details and references are in the text and Online Appendix.\footnote{The lags for the survey variables in \autoref{Model} are implemented by including the auxiliary cycle $\psi^{*j}_{t}$ from \autoref{Cycle}.}

\begin{figure}[t!]
	\centering
	\caption{Prior distributions (in red) and posterior distributions (in blue) of the frequency of the common cycles, persistence of the common cycles, and the variance of the shocks to the common cycles and common trend.}\label{Fig:posterior}
	\includegraphics[width=16cm]{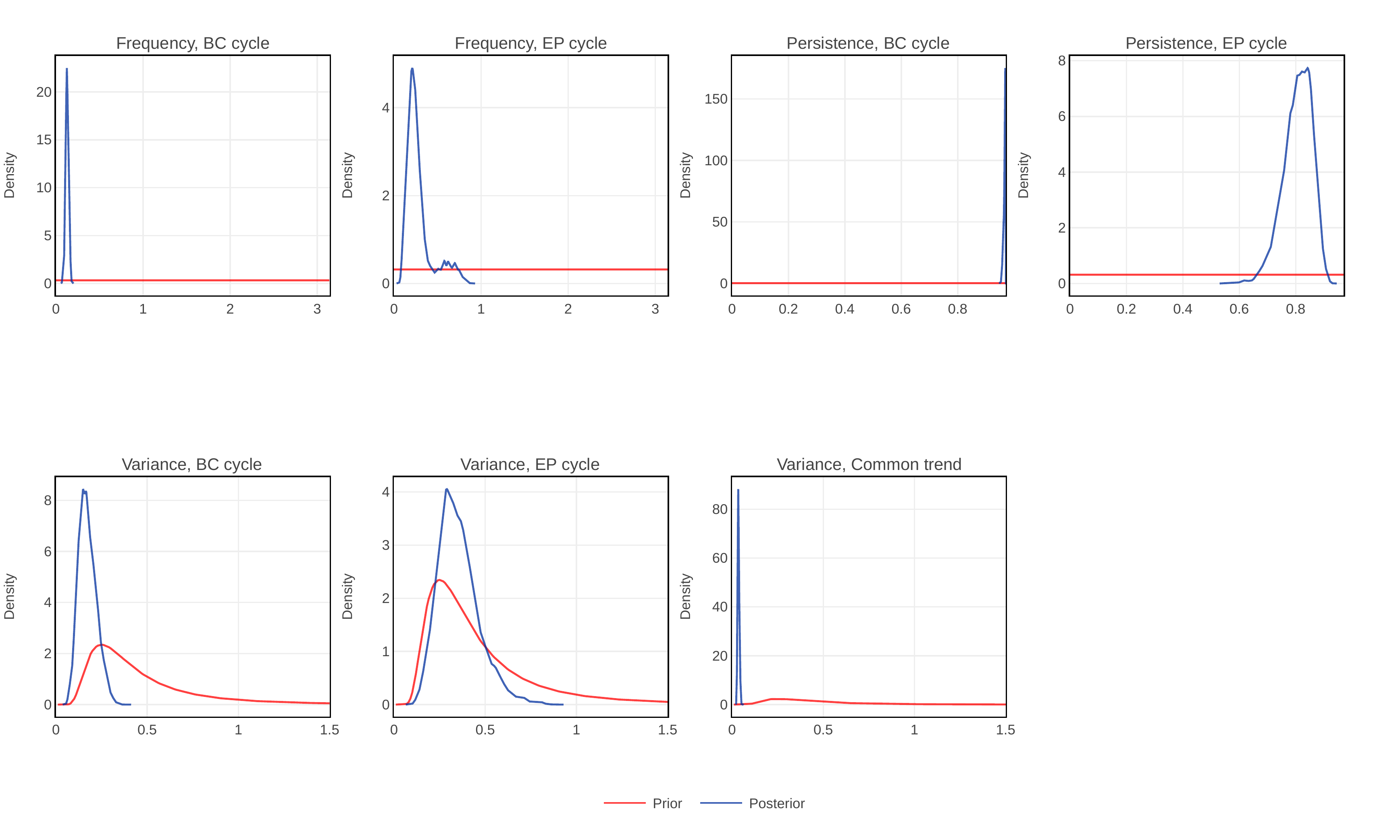}	
\end{figure}

An important question concerns the role of the priors in identifying the model. \autoref{Fig:posterior} and \autoref{Fig:coefficients} illustrate prior and posterior distributions for the variance of the error terms of the unobserved components, the frequency and persistence of the two common cycles, and the coefficients for the common cycles.\footnote{The posterior distributions of the full set of model parameters can be found in the Online Appendix.} 
 The charts provide a good indication on whether data provide enough information to identify the model parameters. Indeed, the posterior distributions are well peaked and not shaped by the priors, and show that the data is very informative in estimating the many parameters of the model -- in particular the variance of the shocks of the common components and the frequencies of the cycles. Importantly, the posterior distributions of the coefficients for the common cycles (\autoref{Fig:coefficients}) indicate that coefficients equal to zero have negligible probability to be drawn in both cases. Moreover, our results are robust to changes in the parameters of the distributions of the more informative priors. See Online Appendix.

\begin{figure}[t!]
\includegraphics[width=12cm]{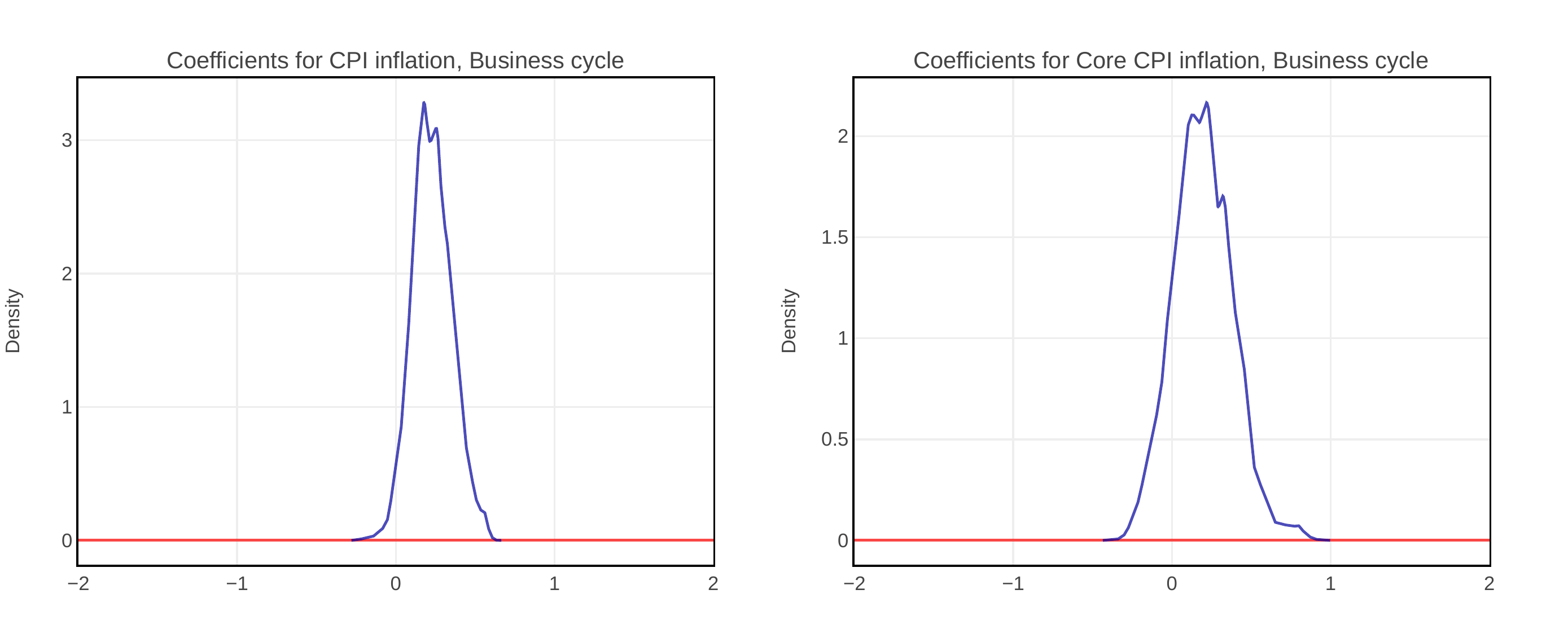}\\
\includegraphics[width=12cm]{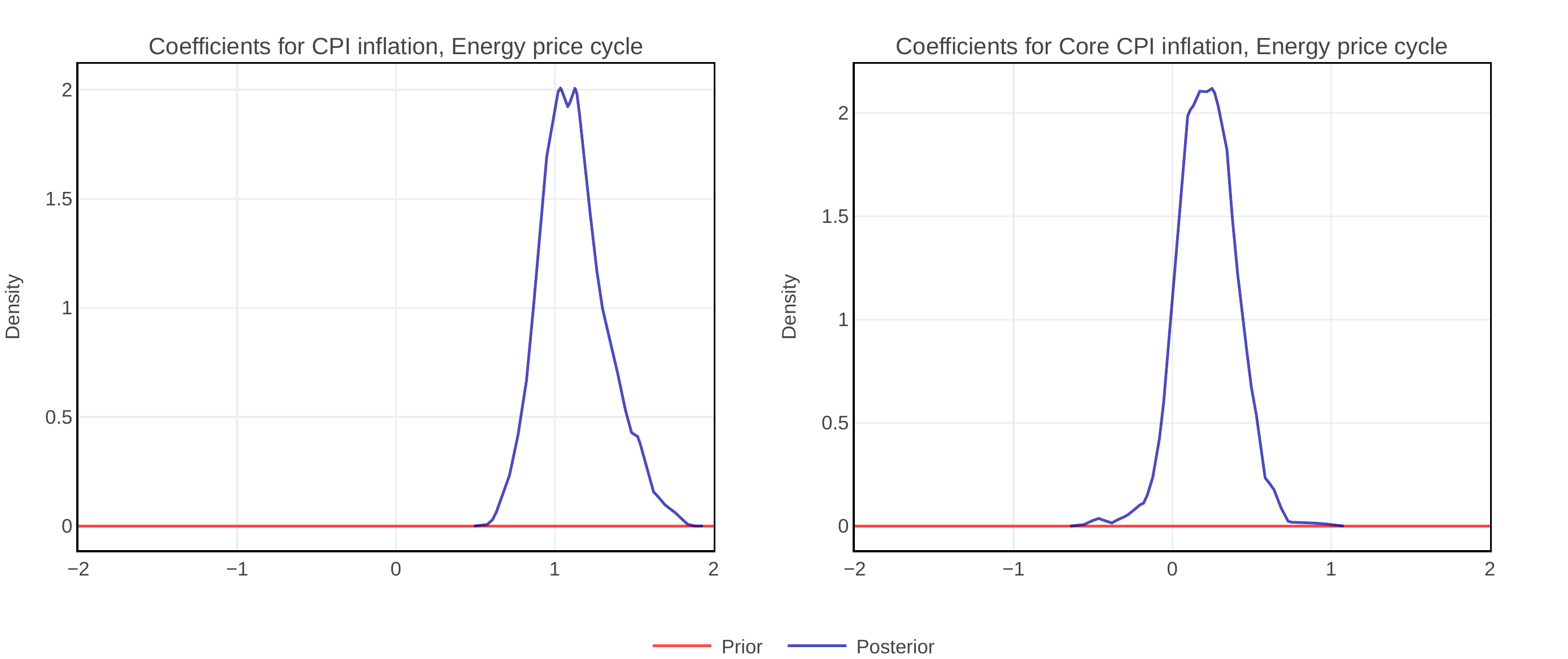}\\	
\caption{Prior distributions (in red) and posterior distributions (in blue) of the coefficients for the common cycles of CPI inflation and Core CPI inflation.}\label{Fig:coefficients}
\end{figure}

\section{Trends and Cycles in the US Economy}\label{sec:empirics}

The empirical model produces a coherent historical narrative of business cycle dynamics and an evaluation of how they impacted  inflation dynamics, as well as a set of model-consistent measures for trend inflation, equilibrium unemployment, and output potential. 

We start by analysing economic trends identified and estimated by the model in the next section and then move to economic cycles in the following one. We compare our assessment of trend-cycle dynamics with the estimates by the Congressional Budget Office (CBO) and the Board of Governors of the Federal Reserve. 

\subsection{Trend Inflation, Equilibrium Unemployment, GDP Potential}

\begin{figure}[t!]
	\centering
	\caption{Independent trends of output, employment, unemployment, and oil prices (in blue), with coverage intervals at 68\% coverage (dark shade) and 90\% coverage (light shade), as estimated by the model. The chart also reports the measures of potential outputs and NAIRU estimated by the CBO (in red).}\label{Fig:trends}
	\includegraphics[width=16cm]{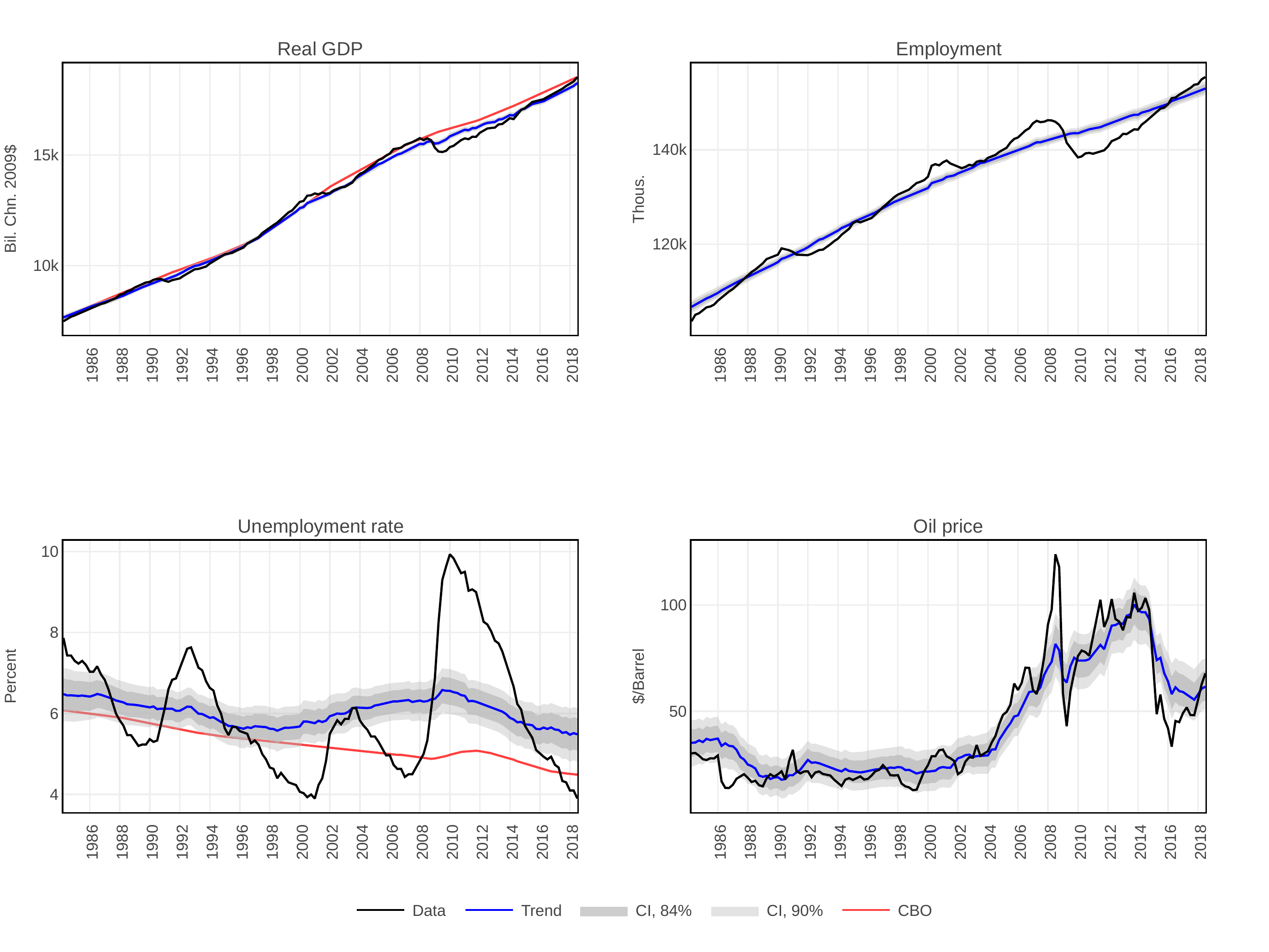}	
\end{figure}

The model delivers very smooth and stable trends. \autoref{Fig:trends} plots real GDP, employment, unemployment, and oil prices against the median of the estimated independent trends, along with coverage bands (at 68\% darker shade, and at  90\% lighter shade coverage rate). Output trend, which can be thought of as a measure of potential output,  is compared  with the corresponding measure provided by the CBO. 

While both trends are equally stable, they provide a different description of long term growth in the US. Since 2001, the model-implied trend lies below the CBO trend implying that, while the CBO's reading of the data is that the US economy had only just reached its potential at the pre-crisis peak in 2008, our model signals an overheating of the economy from 2006 to 2008 and  a marked slow-down of trend growth in the last part of the sample.

\autoref{Fig:trends} also compares the model-implied measure of equilibrium unemployment against the CBO's measure for the natural rate of unemployment (NAIRU). The two measures coincide in the first part of the sample while they diverge post-2000. While our model provides a very stable unemployment trend hoovering around 6\% and with a temporary and small increase around the financial crisis in 2008, the CBO NAIRU shows a slow and persistent decline of the trend continuing through the crisis.\footnote{In the baseline model we include employment measured as number of employed people. However, an important concern relates to the behaviour of the employment-to-population ratio (or active population), which has shown a marked decline since the Great Recession, standing at 61\% in November 2019 down from a pre-crisis level at 63\%. In a robustness exercise reported in the Online Appendix, we substitute employment with employment-to-population ratio in the model. While all of the results reported in this section are robust to the inclusion of this variable, the model captures a persistent decline in the equilibrium trend of the participation rate, following the Great Recession.}

\begin{figure}[t!]
	\centering
	\caption{Trend common to CPI inflation, core CPI inflation, and inflation expectations (in blue), with coverage intervals at 68\% coverage (dark shade) and 90\% coverage (light shade), as estimated by the model.}\label{Fig:common_trend}
	\includegraphics[width=16cm]{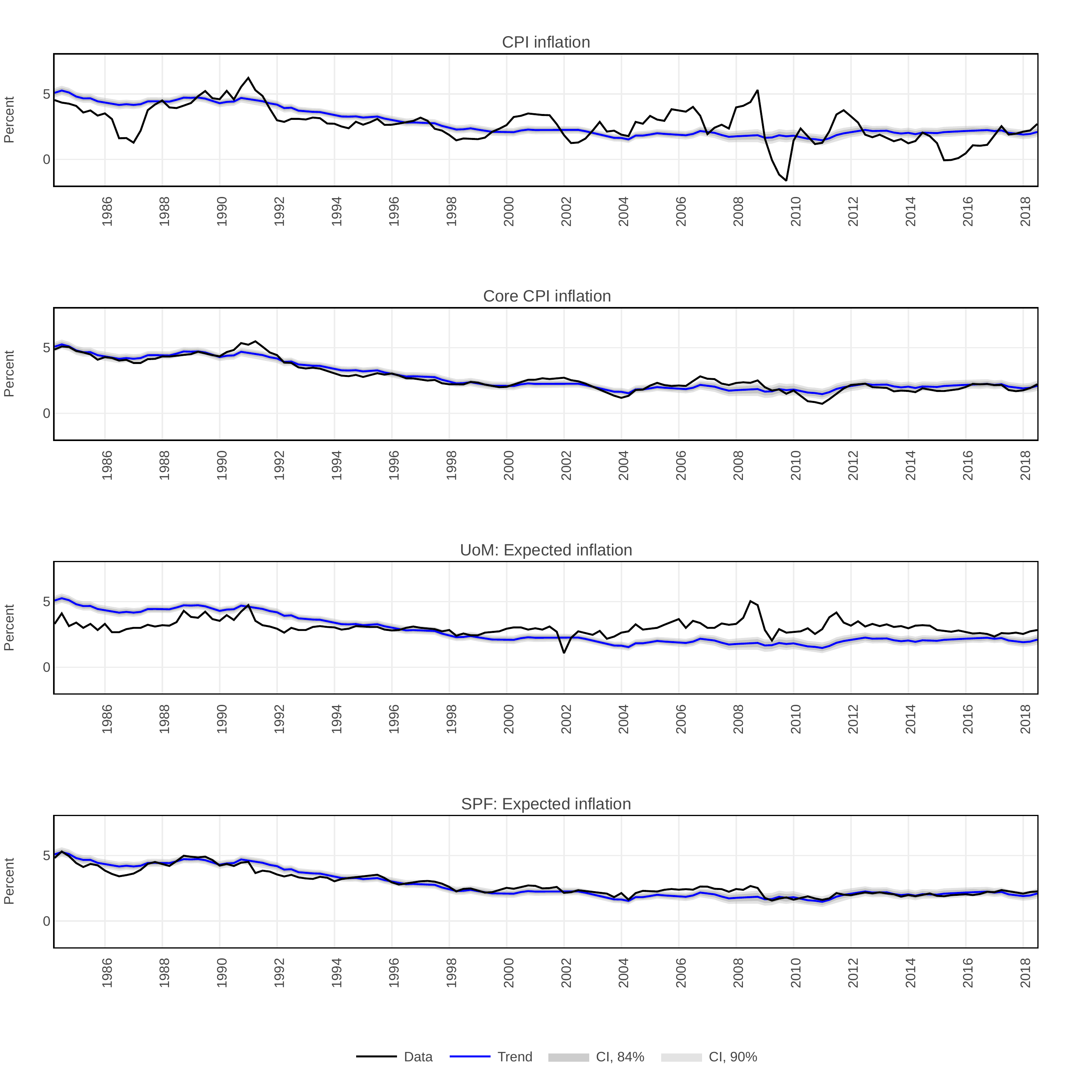}	
\end{figure}
 
The trend in the oil price shows a hump-shaped increase in the second half of the sample that may be related to the global increase in oil demand  post-2000.  It is important to observe that, in our model, trends are  jointly estimated  with the cyclical components. Hence, the differences between our estimated  trends and those of the Fed and the CBO  have relevant implications for the reading of business cycle dynamics. This will be analysed in Section 5.5. 

The inflation trend common to headline CPI, core CPI inflation, and consumers' and professional forecasters' inflation expectation variables is shown in  \autoref{Fig:common_trend}. Trend inflation appears to be roughly stable from 2000 to 2010 and, interestingly, is closely tracked by the SPF median forecast. The behaviour of UoM expectations, on the other hand, shows large and persistent deviations from the common trend (long-term inflation expectations) since 2004. We interpret this sizeable time-varying idiosyncratic trend as a bias in consumers' expectations. 

\begin{figure}[t!]
	\centering
    \caption{Common trend (in blue), with coverage intervals at 68\% coverage (dark shade) and 90\% coverage (light shade), as estimated by the model. The chart also reports the measure of 10 year expectations for CPI inflation from the Survey of Professional Forecasters.}
\label{Fig:common_trend_long_term_exp}
	\includegraphics[width=16cm]{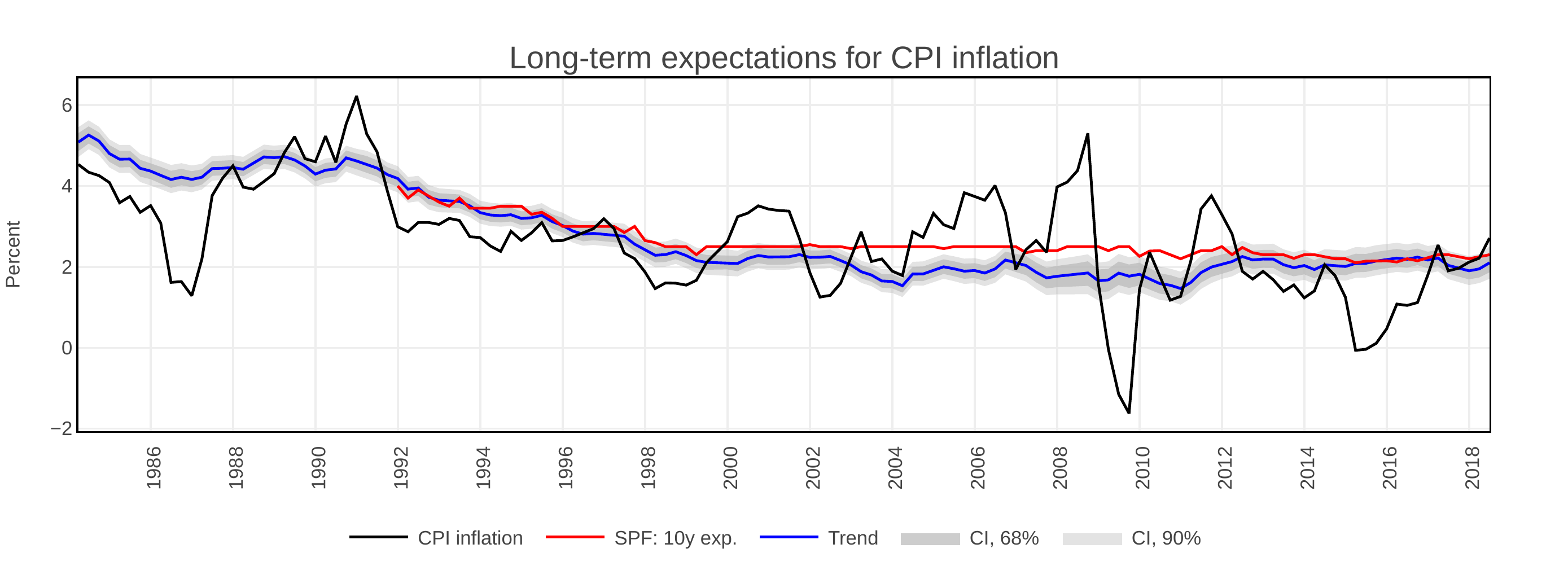}	
\end{figure}

The unit-root inflation trend can be connected to the long-term inflation expectations of rational agents under the assumption that the `law of iterated expectations' holds (see \citealp{beveridge1981new} and \citealp{RePEc:tpr:restat:v:98:y:2016:i:5:p:950-967}). This interpretation is supported by \autoref{Fig:common_trend_long_term_exp} where CPI inflation is plotted against the implied trend and the median 10-year ahead SPF inflation forecast. The chart provides a visual validation of our interpretation that the model trend estimate captures long-term expectations.

\subsection{Business and Energy Price Cycles}

\begin{figure}[t!]
	\centering
	\caption{Top: Business cycle and Energy Price cycle, with coverage intervals at 68\% coverage (dark shade) and 90\% coverage (light shade). Bottom: Parametric spectrum of the Business cycle and Energy Price cycle.}\label{Fig:common_cycle}
	\includegraphics[width=16cm]{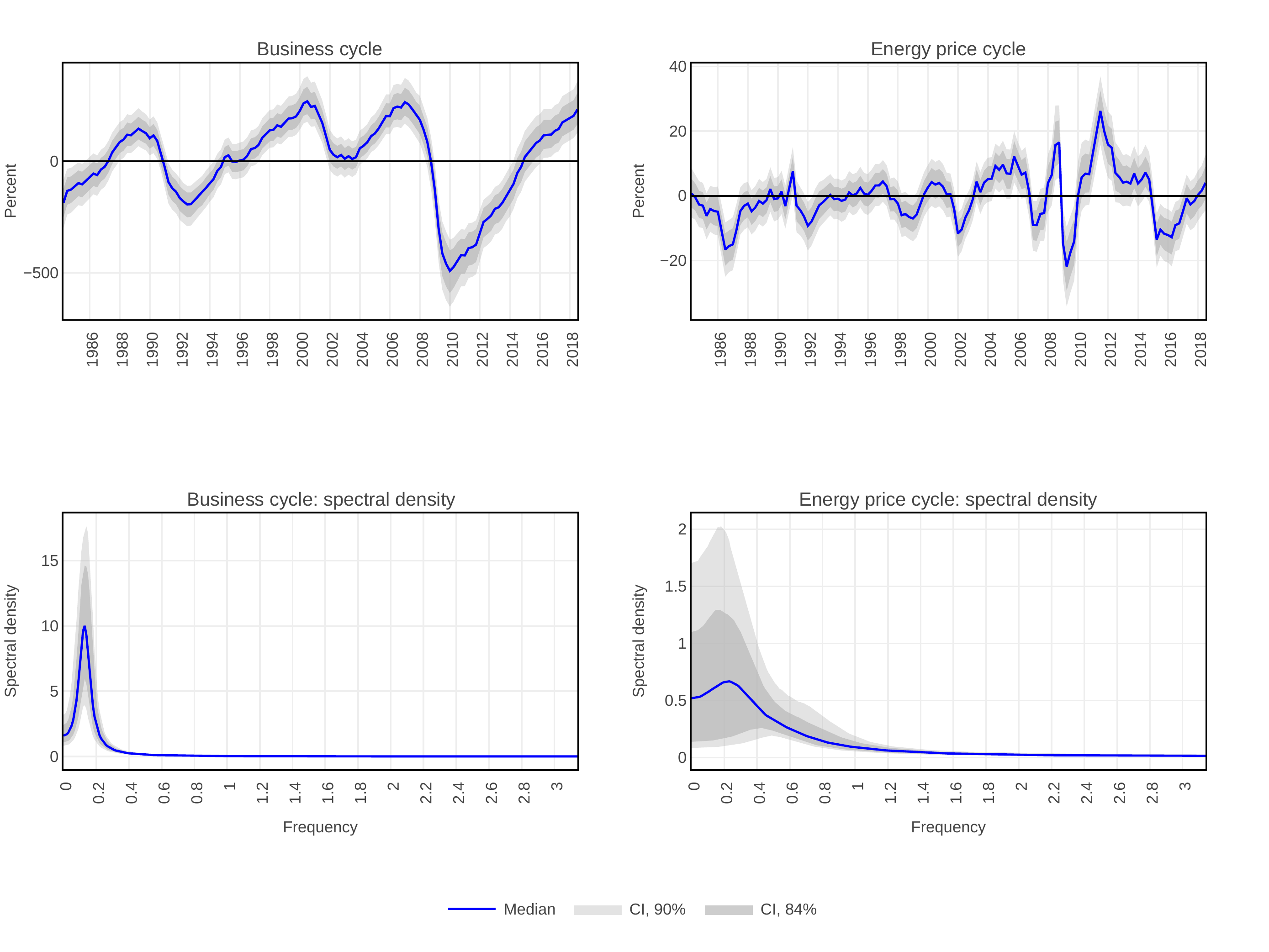}	
\end{figure}

\begin{figure}[t!]
	\centering
	\caption{Top: Business cycle and Energy Price cycle components in CPI inflation, with coverage intervals at 68\% coverage (dark shade) and 90\% coverage (light shade). Bottom: Parametric spectrum of the Business cycle and Energy Price cycle.}\label{Fig:common_cycle_proj}
	\includegraphics[width=16cm]{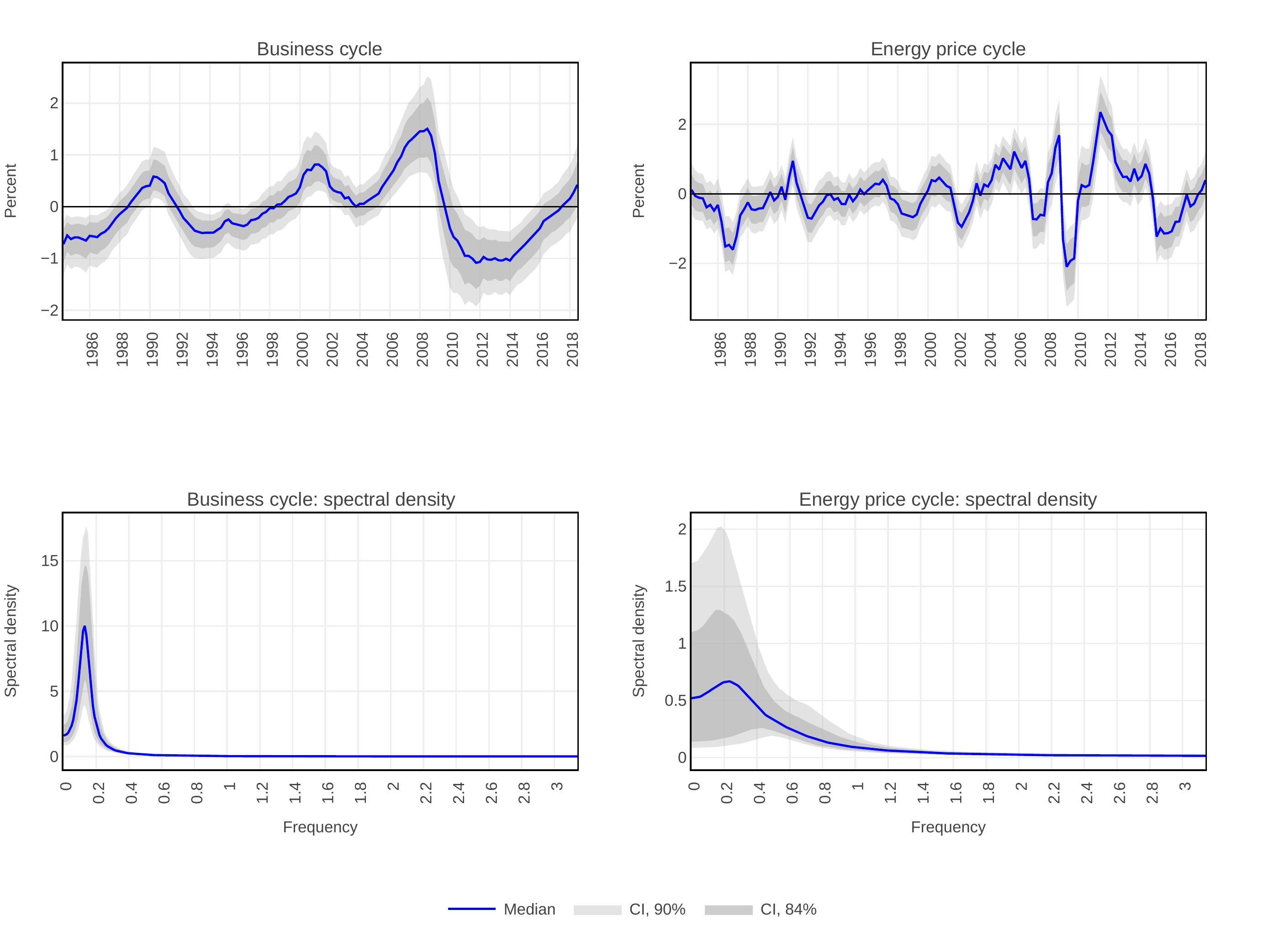}	
\end{figure}

\autoref{Fig:common_cycle} shows the estimated common cycles in both the time and frequency domains, while \autoref{Fig:common_cycle_proj} shows their contribution in headline CPI inflation. The first cycle provides a direct measure of the slack in the economy and captures fluctuations of output around its potential. It also connects real, labour market, and nominal variables and hence can be interpreted as a measure of the business cycle. For this reason, in what follows, we refer to it as `business cycle' with a slight abuse of terminology.  The upper charts in Figures \ref{Fig:common_cycle} and \ref{Fig:common_cycle_proj} report the median of the posterior distribution of the business and energy price cycles with relative coverage intervals at 68\% coverage (dark shade) and 90\% coverage (light shade).  The lower charts show the associated spectral densities and coverage bands. The charts indicate that the `business cycle' is quite regular and much less volatile than the energy price cycle. The spectral shape shows that the business cycle contributes to the inflation spectral shape with a relatively well defined peak and with a cycle between 7 and 8 years periodicity. Conversely, the energy price cycle occupies a broader range of frequencies with a less well defined peak and a periodicity about half as long as that of the business cycle.

\subsection{Historical Decomposition}

\begin{figure}[t!]
	\centering
	\caption{Historical decomposition of the cycles, as estimated by the model. The chart reports the Business cycle (in blue), Energy price cycle (in red), and idiosyncratic cycle (in yellow).}\label{Fig:historical_decomposition}
	\includegraphics[width=16cm]{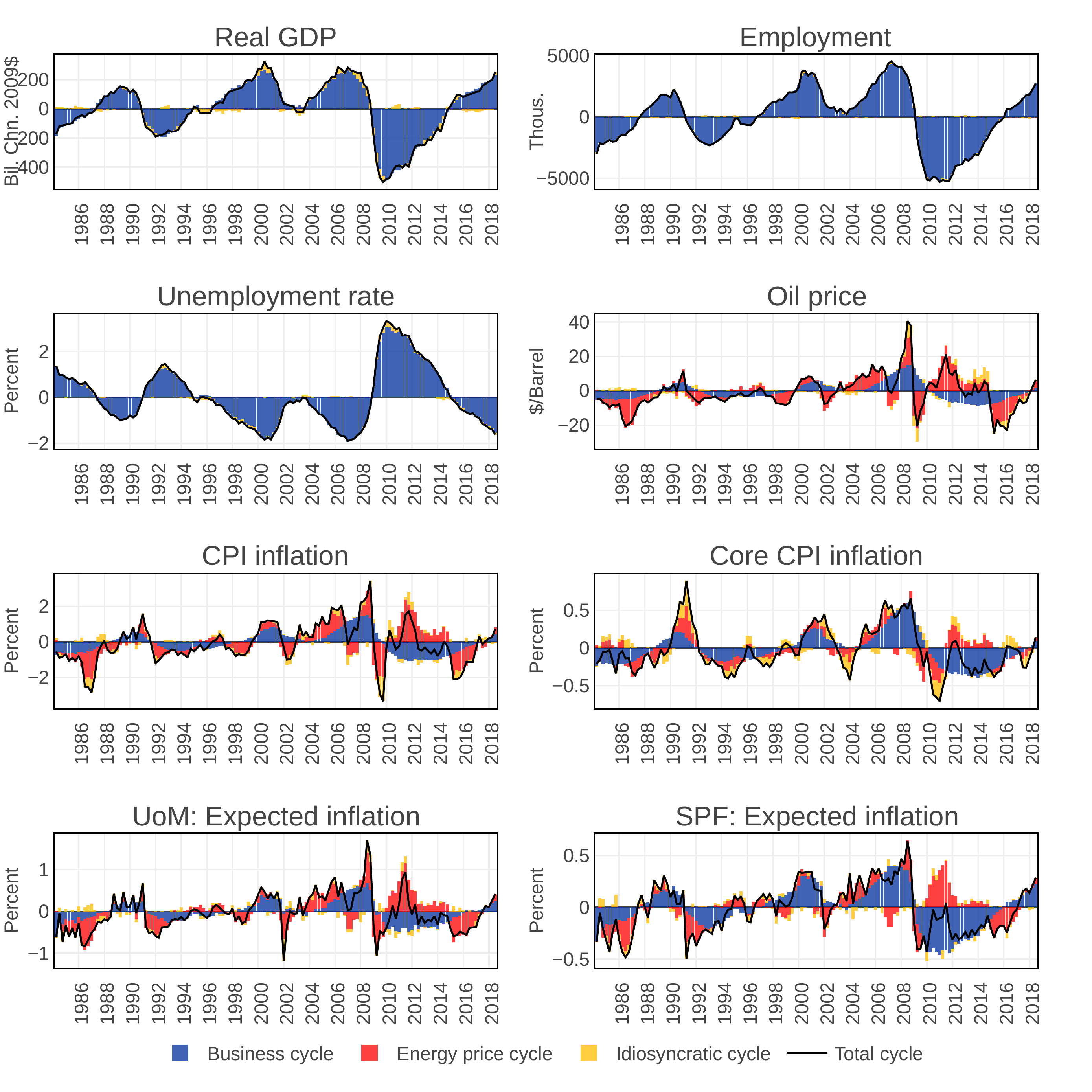}	
\end{figure}

Let us now turn to the historical decomposition of the stationary components of the eight variables of interest into common and idiosyncratic cycles, as provided by the model. \autoref{Fig:historical_decomposition} shows the results. Overall, the model provides a coherent description of inflation dynamics with a number of interesting features. 

First, the business cycle (in blue) capture almost entirely the fluctuations around trend in real output, employment and unemployment. A negligible idiosyncratic component (in yellow) is visible only in unemployment and almost non-existent in output and employment. This indicates that our measure of the output gap captures the slack in the economy well and is transmitted, via the lagged  Okun's law relationship to the labour market. It should be stressed that lags are important in describing the delayed transmission from output dynamics to the labour market and may capture different types of labour market frictions. 

Second, a non negligible share of oil price fluctuations is due to the comovement of this variable with the slack in the economy, along the business cycle. This may be due either to the demand effect of the US economy onto global oil prices, or the role of oil shocks as mark-up shocks in the aggregate production function.

\begin{figure}[t!]
	\centering
	\includegraphics[width=12cm, height=12cm]{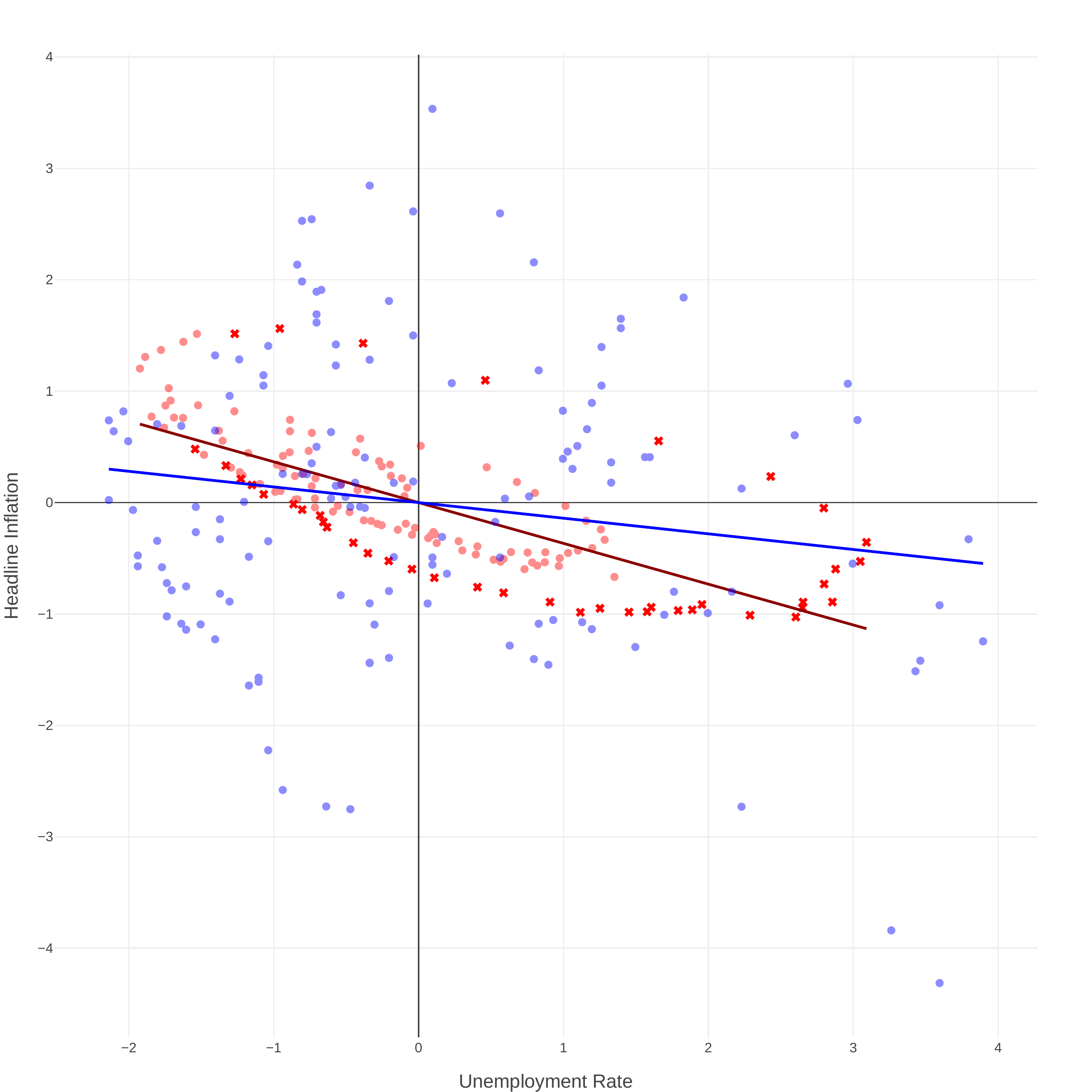}	
	\caption{This chart plots the Business cycle component of CPI inflation against the Business cycle component of the unemployment rate (red dots) and the corresponding bivariate linear regression line (red line). The red crosses represent points the the post Great Recession subsample (from 2008 to 2018). The chart also plots demeaned CPI inflation against the demeaned unemployment rate (blue dots) and the corresponding bivariate linear regression line (blue line).}
	\label{Fig:slope_cpi}
\end{figure}

Third, the slack in the economy is reflected in price dynamics via the Phillips curve which captures the lower frequency dynamics in the inflation cycle and accounts for a sizeable  share of the variation in CPI  inflation and most of the variation in core CPI inflation. This `real' component dominates SPF expectations while it provides a sizeable but not dominant share of variation in consumers' expectations. In our model the Phillips curve is a lagged relationship connecting prices, expectations and output and hence labour market variables, in the spirit of the empirical relationship uncovered by} \cite{ECCA:ECCA283}. A discussion about its `steepness' may be slightly misleading since a reduced form relation between prices and unemployment would involve different lags of our business cycle. Nonetheless, in \autoref{Fig:slope_cpi}, we compare a scatter plot showing how the business cycle components of CPI and unemployment would be related  (red dots) with a   scatter plot of (demeaned) CPI and unemployment variables (blue dots). The linear fit has a slope of -0.39 for the model based measures (red line), against a slope of -0.14 for a na\"ive estimate (blue line).\footnote{The red crosses represent points the the post Great Recession subsample (from 2008 to 2018). Interestingly, the years since the beginning of the last recession seem to be described by the `regular pattern' in the data, albeit they trace a larger than usual `cycle'.} This is a rough way to assess the strength of the Phillips curve identified by our model against that of a na\"ive estimate of its steepness.  

Fourth, the stationary component of CPI inflation is dominated by the energy price cycle. This can be explained by the fact that energy prices are one of the components of the CPI basket and tend to be extremely volatile with a weak correlation with the slack in the national economy. Notice also that, while small, the energy price component is also visible and non-negligible in core CPI inflation where, by construction, energy prices are removed. This  suggests that  oil shocks impact core CPI inflation indirectly via expectations and not via the output gap or other measures of slack in the economy.  In fact, as suggested by \cite{COIBGORO},  household expectations are not fully anchored and respond strongly to oil price changes. Conversely, as observed above, the SPF median forecast tracks the unit-root trends while its cyclical component is dominated by the persistent business cycle component. In other words, the SPF forecasts are  relatively unaffected by the volatile and less persistent energy price component. In this respect, the dynamics of the median SPF forecast seem to be consistent with a rational forecast.

Finally, overall, the cyclical part of inflation is well captured by the two  common components and little is left to idiosyncratic forces. However, the two common cycles are not in any sense `synchronised' . This sheds light on some of the puzzling behaviour of inflation since 2008. From 2011 to mid 2012 the inflation cycle is supported by oil prices while the Phillips curve exerts negative pressure. The opposite is true from 2015 to the end of 2016 when oil prices drag inflation down while the Phillips curve exerts a small upward pressure. 

\subsection{The Role of Oil}

As discussed, oil shocks can impact price dynamics via several different channels. First, as cost-push shocks in production, they impact prices via the Phillips curve. Also, oil prices can fluctuate due to US internal demand along the business cycle. These channels are directly captured by the common business cycle that connects the slack in the economy to oil prices and inflation. Secondly, they directly affect the prices of energy services  which enter the consumption basket of  headline CPI without affecting the output gap. This second channel is likely to explain most of the contribution of the energy price cycle to headline CPI inflation. Thirdly, they can generate `non-fundamental' movement in consumers' inflation expectations and shift prices via this mechanism. This third channel is likely to explain the  energy price cycle component in consumers' expectations and, importantly, in core CPI inflation which excludes  energy prices. Overall, this channel is quantitatively non dominant in price dynamics albeit potentially very important since it is not under the control of standard monetary policy. 

\begin{figure}[t!]
	\centering
	\caption{Idiosyncratic trends of oil prices (left) and UOM Expected Inflation (right), with coverage intervals at 68\% coverage (dark shade) and 90\% coverage (light shade), as estimated by the model.}\label{Fig:idio_trends3}
	\includegraphics[width=16cm]{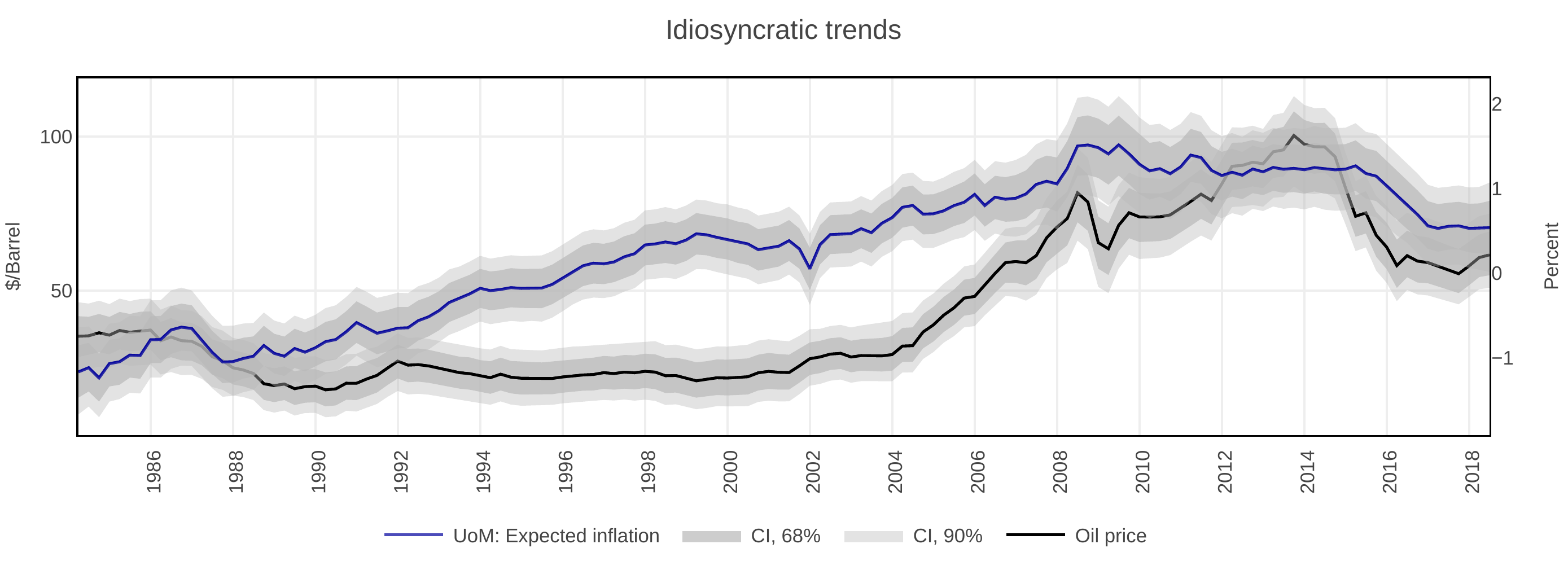}	
\end{figure}

Much of the historical differences in inflation expectations between households and professional forecasters can be accounted for by the contribution of oil prices. This was originally observed by   \cite{COIBGORO} who also attribute to oil shocks a sizeable effect on consumer expectations.  In our framework the effect can only be present through common stationary cycles and trends. However, our results show that there is a large idiosyncratic trend component in oil prices which, by construction, does not affect CPI inflation.   \autoref{Fig:idio_trends3} plots it against the idiosyncratic consumers' expectation trend and  provides suggestive evidence that consumer price expectations may actually have a persistent component related to  oil prices. Our framework leaves it as unmodelled, and to future research. 
 
\subsection{The Output Gap and a Narrative of the Great Recession}

\begin{figure}[t!]
	\centering
	\caption{Output gap (in blue), with coverage intervals at 68\% coverage (dark shade) and 90\% coverage (light shade), as estimated by the model. The model-based estimates of the output gap is obtained by rescaling the business cycle to match the GDP scale and by summing to it the output idiosyncratic cycle component. The chart also reports the output gap from the CBO (in red) and the Fed's Greenbook (in green).}\label{Fig:output_gap}
	\includegraphics[width=16cm]{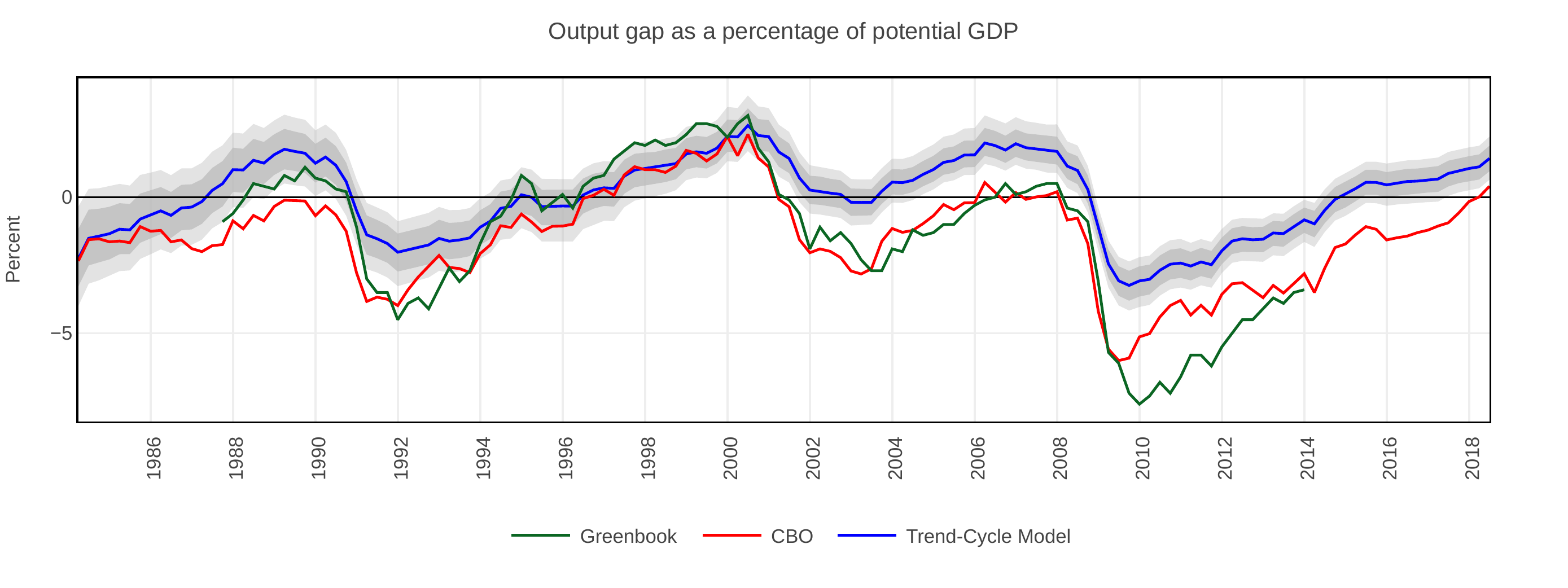}	
\end{figure}

 In the narrative emerging from the model, the output gap has a crucial role. \autoref{Fig:output_gap} reports the model-based output gap as well as the gap published by the CBO and the one by the Fed Greenbooks. The model's and the CBO/Fed business cycle dating of the turning points perfectly coincide as the peaks and troughs alignment shows. However, the model-consistent measure and the other two differ in their assessment of the the degree of slack in the economy since 2001. In fact, at the time of the  slowdown of 2001-2002, our model indicates that the economy went from over-capacity to trend growth but, unlike the CBO's, does not identify a protracted period of slack. 

Notably the model attributes a smaller share of the reduction in GDP following the Great Recession to its cyclical component -- as compared as the CBO's and by the Fed Greenbook's estimates -- and hence projects a lower output potential with a marked slow down in output trend growth that starts before last recession but that becomes manifest in its aftermath (in \autoref{Fig:trends} and \autoref{Fig:output_gap}).  The CBO has a more optimistic assessment of the trend growth and attributes the slowdown since the early millennium to a very deep contraction in the cyclical component of output. Its estimated output gap considers the US economy to have been below potential since 2001 and even at the height of the peak preceding the Great Recession, when the US economy was supported by the unusual dynamics in the real estate market. 

It is important to observe that the two different narratives are the specular image of the question regarding the stability of the Phillips curve. Our model's estimate of the output gap is informed by loose priors on trends, the inflation trends implicit in agents' expectations, and above all the multivariate links connecting prices to the labour market and to output. In doing this, it assumes the stability of the Phillips's curve and of the Okun's Law. It finds that the data matches this description but shows a substantial decline in output potential (and a roughly constant equilibrium unemployment). Conversely, a view of the US economy assuming a very stable potential output implies a widening output gap and hence a flattening of the Phillips Curve. 

Both interpretations are plausible. The two different narratives of the economic developments since 2001 are based on different and untestable assumptions about the long run behaviour of output and other variables and there is no obvious criterion on the basis of which we can choose the `correct' one (see, for example, the discussion on trends in \citealp{RePEc:eee:econom:v:95:y:2000:i:2:p:443-462}). 

Several narratives are compatible with the model's assessment. For example, \cite{Halletal} have pointed to a lower productivity growth trend preceding the Great Recession and, using a growth accounting framework, have argued that the slowdown was due to long-term trend in labour force participation and TFP growth. The slowdown in the pre Great Recession period may have been masked by the dot-com bubble first and the financial boom later, possibly in line with \cite{RePEc:oup:oxecpp:v:69:y:2017:i:3:p:655-677.}. This `productivity view' contrast with an `hysteresis view' on the post-crisis period \citep[see the discussion in][as an example]{RePEc:iie:wpaper:wp15-19}.\footnote{\cite{RePEc:iie:wpaper:wp15-19} using multi country data but not a model based approach conclude that several recessions of different nature are followed by lower growth. They conclude that ``in many cases, the correlation between recessions and subsequent poor economic performance reflects reverse causality: the realization that growth prospects are lower than was previously assumed naturally leads to both a recession and subsequent poor performance.'' However, in other cases ``hysteresis, and perhaps even super-hysteresis may indeed also be at work.''} Deep recessions can cause hysteresis in the form of permanent (or very persistent) changes to trend growth.

It is important to stress, that our model is not able to discriminate between these contrasting narratives and cannot disentangle the source or the nature of this slowdown (see \citealp{RePEc:nbr:nberwo:23580} for a discussion on the issue).\footnote{\cite{RePEc:bin:bpeajo:v:49:y:2018:i:2018-02:p:343-441} observe that ``one should draw little inference from the evolution of estimates of potential GDP about the persistence of GDP changes; these estimates fail to exclusively identify supply shocks that should drive potential GDP and instead also respond to transitory demand shocks. The fact that most of the output declines observed since the Great Recession are now attributed to declines in potential GDP implies little, other than that these declines have been persistent because estimates of potential GDP fail to adequately distinguish between the underlying sources of changes in GDP.''} Indeed, from a statistical perspective, it identifies a very persistent slowdown in output growth that does not fit the usual business cycle dynamics. Hence, it attribute the very persistent reduction in output growth in the aftermath of the Great Recession to changes in the trend growth, and hence possibly to either forms of hysteresis or long-term trends.

\section{Global Factors in US Inflation}\label{sec:global}

\begin{figure}[h!]
	\centering
	\caption{Historical decomposition of the cycles, as estimated by the model. The chart reports the Business cycle (in blue), Energy price cycle (in red), and idiosyncratic cycle (in yellow).}
	\includegraphics[width=14cm]{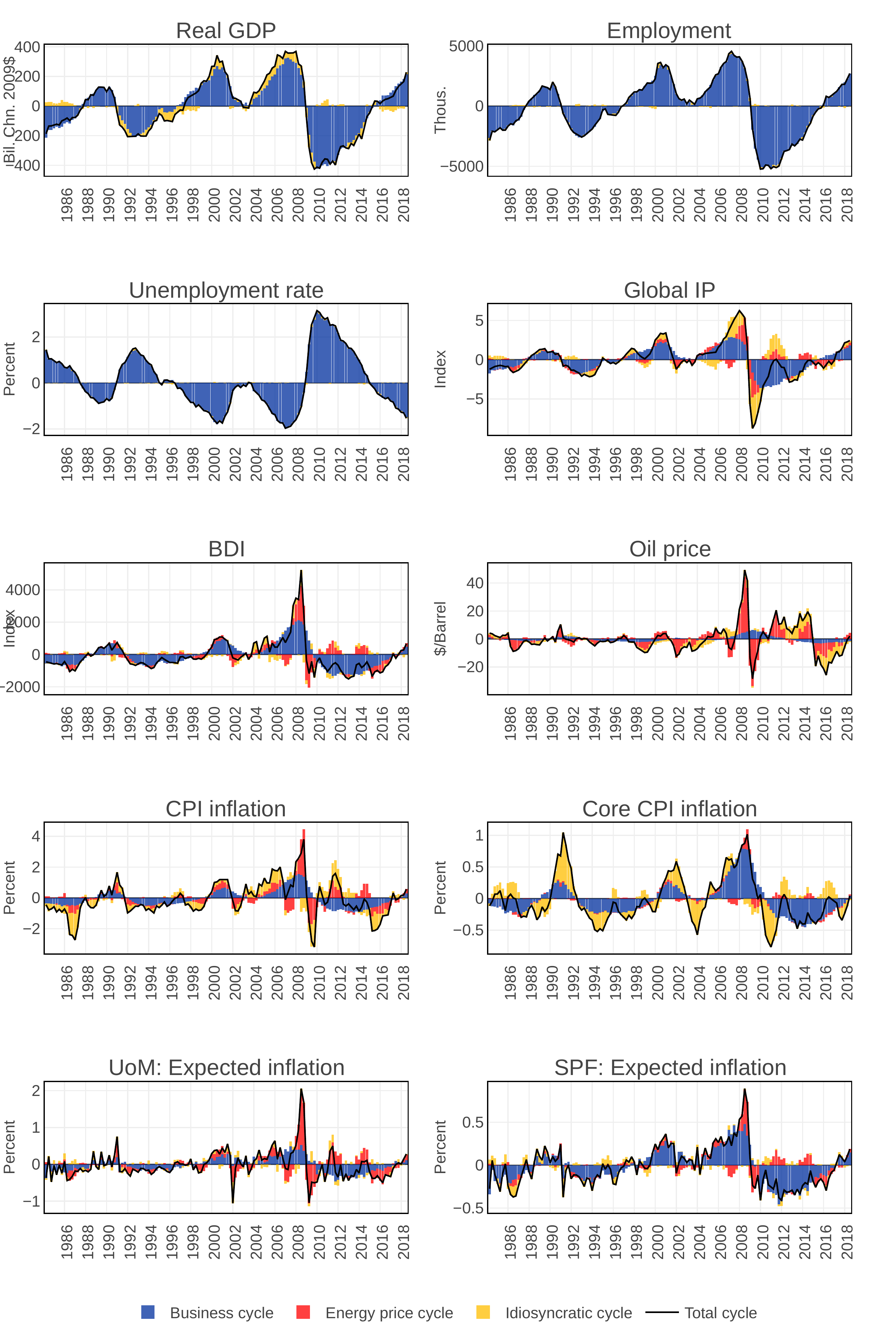}\label{fig:global_cycles}	
\end{figure}

In recent years, the potential impact of globalisation on price dynamics has drawn  attention from both policymakers and academics. The literature has suggested that the increase in international trade has negatively impacted the strength of the domestic Phillips curve relationship and increased the significance of  `global slack' and exchange rates in relation to CPI.  Several channels have been proposed including the increasing impact of demand from emerging markets that has affected volatility in commodity prices, the increased price competition and the greater role of supply chains have reduced  firms' pricing power, or that the reduced bargaining power of local workers has weakened the role for domestic slack \citep[see][for a theory-informed discussion of the literature on the topic]{RePEc:ijc:ijcjou:y:2010:q:1:a:5}. 

Indeed, a number of empirical works have identified a sizeable global common factor in inflation dynamics (e.g. \citealp{doi:10.1162/RESTa00008}, and \citealp{RePEc:red:issued:09-235}), or proposed to add a measure of global slack (e.g. \citealp{RePEc:bis:biswps:227}, \citealp{RePEc:eee:jimfin:v:29:y:2010:i:7:p:1340-1356}), supply chain intensity (e.g. \citealp{RePEc:eee:moneco:v:57:y:2010:i:4:p:491-503, RePEc:snb:snbwpa:2017-03}) or exchange rates (e.g. \citealp{RePEc:mpc:wpaper:0050}) in the econometric specifications of price equations.  

In our analysis we have so far abstracted from these considerations. We instead focussed on the energy price cycle which we extracted   as a process that is orthogonal to domestic slack and not reflected in the output gap and in the labour market conditions in the US. An important question is whether the energy price cycle reflects global demand and  commodity price cycles, as suggested, for example, by \cite{RePEc:ecb:ecbrbu:2018:0051:}. To try and address this question, we estimate a new version of the model that expands the benchmark specification by including the two different measures of global activity: (i) the Baltic Dry Index and index of global cargo shipments, initially proposed by \cite{10.1257/aer.99.3.1053} but taken in levels; (ii) the measure of Global Industrial Production proposed by \cite{baumeister2019structural} and based on the OECD methodology.\footnote{In an explorative analysis reported in the Online Appendix, we provide scatter-plots and correlation coefficients for the business and the energy price cycles in relation to three variables measuring global activity:  (i) the Baltic Dry Index; (ii) the global industrial production (GIP); and (iii) the Global Condition Index (GCI) of \cite{RePEc:fip:fedgin:2018-06-15}.} 

In this new specification, all the variables in the model are allowed to load onto the US business cycle as a reflection of the global significance of the US economy both in terms of share of world GDP and as driver of global economic activity. As in the benchmark specification, US GDP and labour market variables do not load on the energy price cycle, while all the others -- including the Baltic Dry Index and global industrial production -- can have an impact on it.\footnote{The Online Appendix reports details of the model and additional charts.}

In the new specification, the decomposition of the US variables in terms of the BC and the EP is largely unchanged, despite the introduction of global variables, as reported in Figure \ref{fig:global_cycles}. This is reassuring and shows that results are robust. However, the new model offers interesting insights on the role of global shocks in the US inflation dynamics. 

First, the US business cycle drives a large portion of the global economy and hence of the global business cycle fluctuations. This is visible in the large share of the two global indicators explained by the US business cycle component and that is due to both the US weight in world GDP but also to the share of the global activity that is synchronised on the US business cycle.

Second, the energy price cycle now explains a sizeable share of the Baltic Dry Index and oil prices but a smaller share of Global Industrial Production. A possible interpretation is that the fluctuations captured by the energy price cycle are due to oil supply shocks  and possibly financial shocks in the commodity markets, rather than to global demand factors. Interestingly, in the global model, the spectral shape of the energy price cycle is well defined and peaks in a range higher than business cycle frequencies.

\section{Model Forecasting Performance}\label{sec:forecasting}

In the previous sections we showed that a trend-cycle model, incorporating key economic relations and allowing for deviations of agents' forecasts from full information rational expectations, provides a coherent `structural' interpretation of economic developments in the US from the 1980s onwards, based on fundamental and generally accepted economic relationships. While this is an important and desirable feature of the `in-sample' behaviour of the model, an additional test of robustness and reliability of the model is provided by its out-of-sample behaviour. 

In this section we provide an out-of-sample assessment of the model along two dimensions. First we look at trends and cycles extracted by the model in expanding samples, as it would happen in out-of-sample forecast, and check for their stability. This is important in assessing whether the historical decomposition provided by the model is reliable in a pseudo-real-time exercise. Second we test the out-of-sample forecasting performance of the model against two of the best performing models used for inflation forecasting. Forecasting inflation is notoriously difficult and good performance from such a complex model would provide indirect evidence of whether the model is able to capture important features of the data generating process. 

\begin{figure}[ht!]
	\centering
	\caption{This chart shows the revisions of the business cycle (top), energy price cycle (middle), and common trend (bottom) as estimated during the OOS forecasting exercise.}
	\includegraphics[width=16cm]{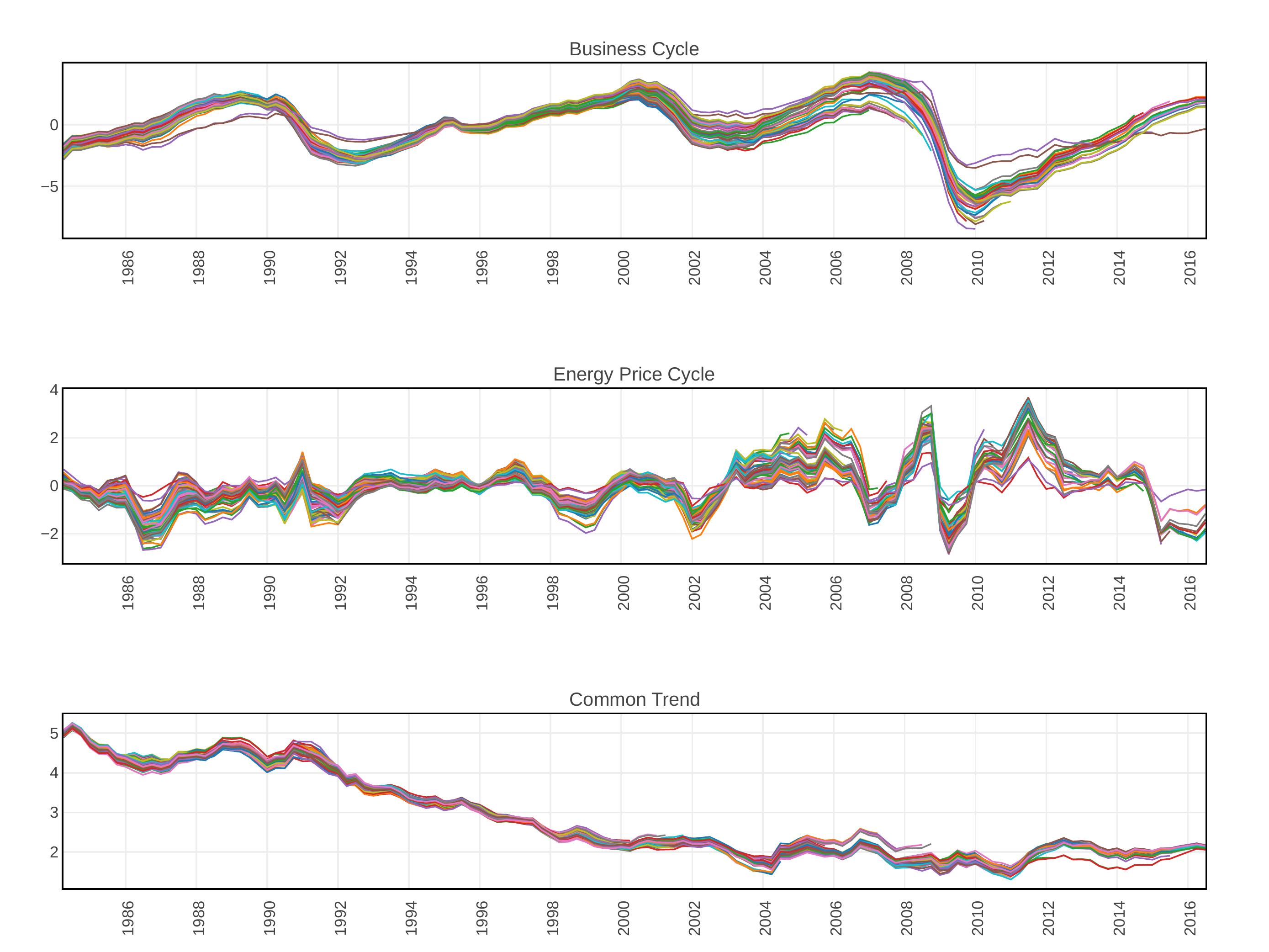}
	\label{Fig:revisions}	
\end{figure}

Figure \ref{Fig:revisions} shows the revisions of the two common cycles and of the inflation trend over time with an expanding data window. The model is re-estimated every quarter. The period from Q1 1984 to Q4 1998 is employed as the pre-sample, while the evaluation sample starts in Q1 1999 and ends in Q2 2017. Results show that trends and the common business cycle are fairly stable overall and provide an assessment of the development in the economy that is evenly consistent over the sample - including in the recessions. The energy price cycle provides a slightly less stable, albeit roughly coherent, reading of the contribution of energy fluctuations to prices. 

The forecasting exercise is conducted in the same sample and again the period from Q1 1984 to Q4 1998 serves as the pre-sample. We use an expanding window and recursively forecast up to 8 quarters ahead. 
In every quarter we reestimate the model, including the unobserved components and the coefficients. Apart from our model (TC), we consider (i) a BVAR where priors are set as in \cite{giannone2015prior}, (ii) a BVAR with ``long-run'' prior as in \cite{giannone2019priors}, and (iii) an univariate unobserved components IMA(1,1) with stochastic volatility model as suggested by \cite{stock2007has} to be tough benchmarks for inflation forecasts. In setting the system with long-run priors we try to closely replicate the main assumptions on trends adopted in our trend-cycle model. In particular, we set long-run priors considering a common trend between CPI inflation, core CPI inflation, and inflation expectations. We allow for the difference between core CPI inflation and CPI inflation, and the difference between inflation expectations and CPI inflation to be stationary.\footnote{In the Online Appendix, we provide details on how the long-run priors are elicited, following an approach that is analogous to the one followed in designing the trend-cycle model.} 

For all models we report the root mean squared forecast errors relative to those of a random walk with drift for forecasting horizons of one, two, four, and eight quarters. We also report the test statistical significance of each benchmark model forecast against the trend-cycle model forecast using \cite{diebold1995} with a quadratic loss function and the modification from \cite{HARVEY1997281}. The Diebold-Mariano test assesses the null hypothesis that the benchmark model forecast and the trend-cycle model forecast are equally accurate. Hence, if the relative root mean squared forecast error of the benchmark model is larger than the relative root mean squared forecast error of the trend-cycle model and we can reject the null hypothesis of equal accuracy, we are allowed to conclude that the trend-cycle model forecast is statistically significantly more accurate than the benchmark model forecast.

\begin{table}[t!]
	\caption{Relative Root Mean Squared Errors}\label{Tab:rmse}
	
	\small
	\begin{tabularx}{\textwidth}{@{}llllll@{}}
		\toprule
		Horizon              & Variable               & TC Model  & MN-SOC-BVAR  & PLR-BVAR & UC-SV  \\
		\midrule
        \multirow{6}{*}{h=1} & Real GDP               & 1.00 & 0.95 & {\bf0.94} &x \\
        & Employment             & 0.94 & 0.76* &{\bf0.75}* &x \\
        & Unemployment rate      & 0.82 & 0.68 &{\bf 0.63} &x \\
        & Oil price              & {\bf1.06} & 1.09 &1.08 &x \\
        & CPI Inflation          & 0.97 & 0.91 &{\bf0.86} &1.00***  \\
        & Core CPI Inflation     & 1.00 & 1.03 &{\bf0.97} &1.01***  \\
        & UOM: Expected inflation& 1.03 & 1.04 &{\bf0.99} &x \\
        & SPF: Expected CPI      & {\bf1.00} & 1.06 &1.06 &x \\
		\midrule                     
        \multirow{6}{*}{h=2} & Real GDP               & 1.02 & {\bf0.96} &0.97 &x \\
        & Employment             & 0.95 & {\bf0.75} &{\bf0.75} &x \\ 
        & Unemployment rate      & 0.80 & 0.72 &{\bf0.65} &x \\
        & Oil price              & {\bf1.08} & 1.18 &1.19 &x \\
        & CPI Inflation          & 0.95 & 0.97 &{\bf0.92} &0.99***  \\
        & Core CPI Inflation     & {\bf0.95} & 1.13 &1.04 &0.99***  \\
        & UOM: Expected inflation& {\bf1.01} & 1.09 &1.04 &x \\
        & SPF: Expected CPI      & {\bf0.97} & 1.18** &1.24* &x \\
		\midrule                     
        \multirow{6}{*}{h=4} & Real GDP               & 1.04 & {\bf1.04}  &1.04 &x \\
        & Employment             & 0.99 & {\bf0.82}  &0.81 &x \\ 
        & Unemployment rate      & {\bf0.81} & 0.84 &0.75 &x \\
		& Oil price              & {\bf1.12} & 1.26 &1.26&x \\
        & CPI Inflation          & {\bf0.95} & 1.12 &1.05 &0.98**  \\
        & Core CPI Inflation     & {\bf0.89} & 1.22* &1.12 &0.96***  \\
        & UOM: Expected inflation& {\bf1.11} & 1.15 &1.10 &x \\
        & SPF: Expected CPI      & {\bf0.91} & 1.28* &1.42** &x \\
		\midrule                     
        \multirow{6}{*}{h=8} & Real GDP               & {\bf1.11} & 1.21 &1.16 &x \\
        & Employment                                  & 1.07 & 1.01 &{\bf0.95} &x \\ 
        & Unemployment rate      & {\bf0.81} & 1.02*** &0.85 &x \\
        & Oil price              & {\bf1.10} & 1.34 &1.35 &x \\
        & CPI Inflation          & {\bf0.85} & 1.07 &0.95 &0.96*  \\
        & Core CPI Inflation     & {\bf0.83} & 1.30** &1.13* &0.91  \\
        & UOM: Expected inflation& {\bf1.02} & 1.29 &1.16 &x \\
        & SPF: Expected CPI      & {\bf0.86} & 1.33* &1.31** &x \\
		\bottomrule
	\end{tabularx}   
    \floatfoot{\textbf{Note:} This table shows the RMSEs relative to the random walk with drift. The MN-SOC-BVAR is a BVAR with ``Minnesota'' and ``Sum-of-coefficients'' priors and was estimated using \cite{giannone2015prior}. The PLR-BVAR is a BVAR with ``long-run prior'' as in \cite{giannone2019priors}. The UC-SV model was first proposed in \cite{stock2007has}. We test that the forecasts of each other model are statistically different from the trend-cycle model forecasts using} \cite{diebold1995} with a quadratic loss function and the modification from \cite{HARVEY1997281}. *p <0.1, **p<0.05, ***p<0.01.
     
\end{table}

Results are reported in Table \ref{Tab:rmse}. They  show that the trend-cycle model outperforms all others for CPI inflation and core CPI inflation at the 4 and 8 quarters ahead horizons. Our conjecture is that our advantage with respect to the two BVARs is driven by the random walk trend which captures the slow-moving, low frequency component. This is consistent with the fact that the advantage of the trend-cycle model over the BVARs is statistically significant for core inflation at least at the 10\% level but not for CPI inflation, since the inflation trend explains a larger fraction of core inflation than of CPI inflation. The advantage of the trend-cycle model with respect to the UC-SV models is most likely due to the Phillips curve which captures cyclical co-movements. This explains why the advantage is more significant at shorter horizons, where the cyclical components in the forecast are larger than at long horizons. The trend-cycle model and the BVARs have similar performance in relation to  the other variables with the exception of employment one quarter ahead where both BVARs outperforms our model with a difference which is statistically significant at the 10\% level. 

Results seem to indicate that despite the large number of parameters and the imposition on the data of structural relationships dictated by economic theory, the model provides a stable historical decomposition in a pseudo real-time exercise and very good performance in forecasting.  We consider this as evidence providing support to the claim that the model is able to capture important features of the data generating process.

\section{Concluding Comments}

The paper proposes a medium size semi-structural model aimed at identifying and estimating the key building blocks of inflation dynamics -- the Phillips curve, long-term inflation expectations, output gap and GDP trend, Okun's law and a high frequency oil price cycle -- within a unified framework.

Results point to a well identified and steep Phillips curve relationship in reduced form, which captures a cyclical component CPI inflation with maximum power at around eight years periodicity but also point to deviations from the standard rational expectations formulation since we identify a sizeable cycle in CPI inflation which is unrelated to real domestic variables and captures the correlation between inflation expectations and oil prices. This cycle, which is of slightly shorter periodicity than the business cycle and  is more volatile, points to a channel through which oil price developments  temporarily affect consumer price expectations away from the nominal-real relationship captured by the Phillips curve. In the presence of large oil price shocks this component may dominate and cloud the signal on cyclical inflation.  The energy price component appears to be determined by global factors such as  oil supply shocks and financial shocks in the commodity markets. 

Interestingly, this energy price cycle is associated to both core and CPI inflation which suggests that even core inflation provides a clouded signal of fundamental (trend and cyclically driven)  inflationary pressures. This result provides motivation to the signal extraction approach we have proposed for the identifiation of the cyclical component of inflation. As for the real variables, the model's estimate of potential output identifies a slowdown around the beginning of the millennium that becomes more evident in the wake of the Great Recession. Our results are compatible with both the  `productivity view' of  \cite{Halletal} and the  `hysteresis view' of \cite{RePEc:oup:oxecpp:v:69:y:2017:i:3:p:655-677.}. The implication is that our estimate of the output gap differs from that of the CBO's  since the beginning of the productivity slow-down. While the CBO's view is that the US economy was growing  around potential before the 2008 crisis and below it since then,  our model points to growth above potential between 2006 and 2008  and again since 2015. 

Although it is not possible to discriminate between these different views that ultimately depend on different beliefs on the long-run behaviour of output,  our model -- based on the joint analysis of output, labor market, prices and expectations -- provides a plausible narrative which is consistent with the data and that can be interpreted in a  transparent way. We believe that as such it provides a useful model-consistent benchmark for the policy debate.

From the policy perspective, our findings indicate that the central bank can exploit the Phillips curve trade-off but only in a limited way since the latter, although well identified, is not the unique determinant of inflation dynamics. Indeed, some of the so-called puzzles of inflation behaviour in the last decade can be explained by disentangling the Phillips curve from the energy price cycle. Moreover, while trend inflation appears to be roughly stable from 2000 to 2010, the behaviour of UoM expectations  shows large and persistent deviations from the common trend (long-term inflation expectations) since 2004 which can be interpreted as a bias in consumers' expectations. Therefore, a problematic issue for the central bank is that, facing volatile and persistent oil price dynamics, consumer expectations can deviate from a stable trend and affect price dynamics. Our conclusions are therefore quite open-ended. The Fed's view that inflation is dominated by three components is supported by the data. However, the ability of the Central Bank to anchor expectations is limited especially because oil affects consumer expectations persistently and independently from the state of the real economy.

\setstretch{1}

\bibliographystyle{aer}
\bibliography{FedInflationModel}
 
\end{document}